\documentclass[unnumsec,webpdf,modern,large]{oup-authoring-template}%

\graphicspath{{Fig/}}

\theoremstyle{thmstyleone}%
\theoremstyle{thmstyletwo}%
\theoremstyle{thmstylethree}%

\usepackage{tabularx}
\usepackage{siunitx}
\usepackage{setspace}
\usepackage{rotating}

\usepackage{float}
\setcitestyle{square}
\usepackage{placeins}

\begin{document}
	
	\journaltitle{PNAS Nexus}
	\DOI{DOI added during production}
	\copyrightyear{2026}
	\pubyear{2026}
	\vol{XX}
	\issue{x}
	\access{Published: Date added during production}
	\appnotes{Paper}
	
	\firstpage{1}
	
	
	\title[Short article title]{From wake dynamics to energy consumption in free-swimming biohybrid robotic jellyfish: a multiscale analysis}
	
	\author[1]{Simon Anuszczyk\ORCID{0009-0000-9840-9442}}
	\author[1]{Kyra Phaychanpheng}
	\author[1,2,$\ast$]{John O. Dabiri}
	
	\address[1]{\orgdiv{Graduate Aerospace Laboratories}, \orgname{California Institute of Technology}, \orgaddress{\street{1200 E California Blvd}, \postcode{91125}, \state{CA}, \country{USA}}}
	\address[2]{\orgdiv{Mechanical and Civil Engineering}, \orgname{California Institute of Technology}, \orgaddress{\street{1200 E California Blvd}, \postcode{91125}, \state{CA}, \country{USA}}}

	\corresp[$\ast$]{To whom correspondence should be addressed: \href{email:email-id.com}{jodabiri@caltech.edu}}
	
	\received{Date}{0}{Year}
	\revised{Date}{0}{Year}
	\accepted{Date}{0}{Year}
	
	
	\abstract{Measuring energy consumption of marine organisms often requires enclosing the animal in a comparatively small, sealed chamber to quantify changes in oxygen concentration of the surrounding water. This can limit measurements of free-swimming organisms by introducing recirculation effects and movement restrictions. We experimentally investigate free-swimming jellyfish energy consumption at two scales: individual pulses and multi-day swimming. Prescribing pulse frequency using onboard microelectronic swim controllers enables the comparison of wake energetics at different swimming stroke frequencies, while also enabling continuous swimming. On the microscale, we quantified pulse wake hydrodynamics using three-dimensional, full velocity field Particle Image Velocimetry. We found electrical stimulation increased posterior wake energy loss 2.9 times compared to unstimulated jellyfish due to heightened pulse rates and modified swimming kinematics.On the macroscale, we used a 6-meter tall, 13,600 liter water tank and animal tracking-based feedback pump control to enable continuous swimming against a flow current without encountering the vertical limits of the tank over 2.55 km. We utilized a non-invasive technique for quantifying changes in 3D morphological reconstructions of the animal without feeding. Changes in animal volume were converted to energy consumption using the body chemical composition determined with elemental analysis. We found free-swimming, electrically stimulated animals consumed 2.5 times more energy than similarly stimulated animals in a constrained environment, consistent with combined hydrodynamic and behavioral differences between free-swimming and enclosed configurations, including increased swimming speed and reduced boundary effects. These results suggest that the observed impact of hydrodynamic drag may be underrepresented in studies relying on confined experimental configurations.
	} 
	
	\keywords{Jellyfish, swimming kinematics, physiology, bioinspired}
	
	\otherabstract[Significance statement]{This paper shows that \textit{Aurelia aurita} jellyfish consume significantly more energy when free-swimming compared to measurements in constrained enclosures where movement is restricted. We externally fix pulse rate to enable continuous swimming and show that energy consumption scales approximately linearly with pulse rate. These results suggest that hydrodynamic and behavioral factors associated with free-swimming conditions may substantially increase metabolic cost relative to enclosed measurements and be underreported in the total metabolism literature of other marine species due to experimental constraints.}
	
	\maketitle
	
	\section{Introduction}
	Energy sourcing, consumption, and allocation are vital to maintaining life. Measuring energy consumption enables the prediction of an organism’s performance and contributes to the characterization of survival, fitness, and phenology \citep{degrootIncorporatingOtolithisotopeInferred2024}. Thus, accurate energy consumption measurements are important not only for understanding physiology but also organismal biology more broadly. 
	
	Locomotion in fluids arises from coupled interactions between animal biomechanics and fluid dynamics. These interactions have long been proposed as a major evolutionary driver shaping the morphology and locomotion of both aquatic and aerial organisms \citep{vogelLifeMovingFluids1994,dabiri_wake-based_2010}. Propulsive forces are generated at the solid-fluid interface, where muscular deformation of fins, bodies, and bells accelerate surrounding fluid \citep{vogelLifeMovingFluids1994}. By conservation of momentum, the impulse imparted to the wake balances the thrust experienced by the swimmer. Thus, animal energetics and performance cannot be understood independently of the surrounding flow field \citep{lighthillMathematicalBiofluiddynamics1997}.
	
Despite advances in both fluid measurements and metabolic techniques, a key gap remains: the relationship between the energy imparted to the wake and the total metabolic cost of swimming in free-swimming organisms is not well understood. Bridging this gap requires simultaneous characterization of wake energetics and whole-animal energy consumption across timescales. 
	
Here, we address this gap by quantifying jellyfish energetics across two complementary timescales, examining the relationship between energy imparted to the wake during individual swimming pulses and total metabolic consumption during prolonged free-swimming. On the microscale, we use volumetric flow measurements to quantify wake energetics, while on the macroscale we apply non-invasive metabolic estimates to measure long-term energy consumption in free-swimming \textit{Aurelia aurita} (Linnaeus, 1758).
	
On the microscale, despite the central role of wake dynamics in propulsion, most experimental measurements of swimming flows have historically been restricted to two-dimensional velocity fields \citep{costello_hydrodynamics_2021,elsingaTomographicParticleImage2006}. Planar Particle Image Velocimetry (PIV) captures velocity components within a single measurement plane, which can obscure out-of-plane transport of momentum and energy associated with inherently three-dimensional wake structures such as vortex rings \citep{adrianParticleImageVelocimetry2011}. This limitation is particularly important for many medusae whose propulsion is dominated by the formation and evolution of vortex rings shed during bell contraction and relaxation \citep{dabiri_wake-based_2010,costello_hydrodynamics_2021}. As a result, two-dimensional measurements may underrepresent the full spatial distribution of wake momentum and kinetic energy generated by swimming jellyfish.
	
Recent advances now enable simultaneous measurements of three-dimensional velocity fields and animal biomechanics. Multi-camera systems such as stereo PIV extend traditional two-dimensional two-component PIV to enable the capture of the out-of-plane velocity component for a single two-dimensional plane \citep{adrianParticleImageVelocimetry2011,elsingaTomographicParticleImage2006,fu_single-camera_2021}. Tomographic PIV expands this system to enable the capture of full volumetric velocity fields in three dimensions. These systems can be cumbersome to set up and calibrate due to significant camera and equipment requirements, which often limit experimental volume \citep{elsingaTomographicParticleImage2006}. A similar system was previously used on \textit{Aurelia aurita} vortex rings with small radii of a few cm \citep{gemmell_control_2015}. 
	
Here we utilize a single camera volumetric PIV scanning system to improve on these limitations while recording full velocity fields in three dimensions \citep{fu_single-camera_2021,mohebbiMeasurementsModellingInduced2024}. This technique enables direct quantification of wake dynamics, vortex structures, and kinetic energy, thus providing a more complete framework for linking kinematics to energetic cost.
	
On the macroscale, a variety of techniques have been developed to measure the metabolism of marine organisms. The most common of these techniques, respirometry, measures the respiratory rate, which is a metabolic measurement of oxygen consumption commonly used as a proxy for energy consumption. Respirometry comprises three main techniques, each with associated benefits and drawbacks \citep{svendsenDesignSetupIntermittentflow2016}. Closed system respirometry encloses the animal in a tank, generally of a similar size as the animal, and measures the decreasing oxygen concentration in the water as the animal respires. These systems are simple and ubiquitous, but avoiding hypoxic effects and the buildup of organic waste in the chamber limits experimental duration. Flow-through respirometry instead provides a continuous flow of water through the chamber and measures the difference in oxygen concentration at the inlet and outlet to calculate oxygen consumption. While this technique minimizes hypoxic effects, concerns of sensor issues, mixing, and waste buildup due to imperfect mixing persist. Finally, intermittent-flow respirometry combines periods of closed system respirometry with flushes of the chamber to eliminate both hypoxic effects and the buildup of waste \citep{svendsenDesignSetupIntermittentflow2016}. However, all of these methods rely on changes in oxygen consumption measurable above background noise. In practice, each requires comparatively small experimental chambers, which introduce other confounding variables such as movement restriction and accompanying animal stress \citep{harrisICESZooplanktonMethodology2000}.
	
Other techniques attempt to avoid the limitations of small experimental chambers by studying animal energy consumption in situ. Biotelemetry, including electromyogram sensors and accelerometry, has been used to indirectly estimate energy consumption. These techniques are most applicable to large fish due to tag size and generally do not control for different types of activity. Like doubly labeled water, isotopic trace turnover methods such as carbon isotopes consumed through feeding are valuable for energy consumption measurements. However, these techniques require extensive validation before application to measure decay curves for each species \citep{trebergEstimatesMetabolicRate2016}. Although these techniques enable energetic measurements in more realistic free-swimming environments, their accompanying limitations make them unsuitable for gelatinous zooplankton.
	
Here, we develop, characterize, and implement a non-invasive technique that overcomes these limitations by inferring the energy consumption of free-swimming \textit{Aurelia aurita} jellyfish medusae based on changes in body volume due to catabolysis. Catabolysis is the process by which an organism breaks down its own tissue when deprived of nutrients. Decreases in size during periods of catabolysis have been documented in many marine organisms, including fishes \citep{simpkinsPhysiologicalResponsesJuvenile2003}, jellyfish \citep{lilleyIndividualShrinkingEnhance2014,ishiiFoodRegulationGrowth1998,fuBodySizeReduction2014}, and other organisms. Previous work has shown starving \textit{Aurelia aurita} jellyfish mass decreases by up to 13.4\% per day \citep{ishiiFoodRegulationGrowth1998}, and some work predicts a derived respiratory carbon demand of more than 17\% per day \citep{bondyale-juez_wind_2022}. One study inferred metabolic rates from average mass changes of \textit{Aurelia aurita} ephyrae to estimate the respirometry rate. However, this study relied on euthanasia as part of their protocol for measuring animal mass, limiting the applicability of this technique \citep{fuBodySizeReduction2014}. Recent developments in volumetric (3D) laser scanning enable detailed 3D reconstructions of gelatinous zooplankton \citep{fu_single-camera_2021,danielsNewMethodRapid2023}. This allows for measurements of changing body size due to catabolysis while in situ, which can be converted to energetic consumption rates \citep{harrisICESZooplanktonMethodology1999}.
	
	\subsection{Electrical stimulation}
Jellyfish pulse rate, and thus metabolism, varies significantly based on environmental conditions and animal morphology \citep{tillsReducedPHAffects2016,dillonEffectsAcuteChanges1977,mchenryOntogeneticScalingHydrodynamics2003}. To systematically investigate the relationship between wake energetics and metabolic cost, we used onboard electrical stimulation to fix animal pulse rate at a defined frequency across animals and throughout the duration of the experiments, thus creating a biohybrid robotic jellyfish \citep{xu_low-power_2020}.
	
\subsection{Experimental plan}
On the microscale, we measured the flow around swimming jellyfish on the order of individual pulses. We conducted three-dimensional, three-component PIV on free-swimming jellyfish. Comparison of unstimulated and electrically stimulated animals elucidated differences between the induced flows and biomechanics of these jellyfish.
	
On the macroscale, we measure volumetric changes in free-swimming electrically stimulated jellyfish. Testing over swimming durations greater than 15,000 body lengths enabled assessment of long-term changes in metabolism during catabolysis. We compared these measurements against predictions from a physics-based energetics model and basal metabolism experiments.

\section{Materials and Methods}
We conducted experiments to quantify jellyfish energetics across microscale wake dynamics and macroscale metabolic consumption.

\subsubsection{Animal husbandry}
\textit{Aurelia aurita} medusae were obtained from Cabrillo Marine Aquarium (San Pedro, CA, USA) and Aquarium of the Pacific (Long Beach, CA) and housed in a 453 liter pseudokreisel tank. See previous work for more details \citep{anuszczykElectromechanicalEnhancementLive2024}.
	
\subsubsection{Experimental manipulation using electrical stimulation}
	
	\begin{figure}[!htbp]
		\includegraphics[width=\columnwidth]{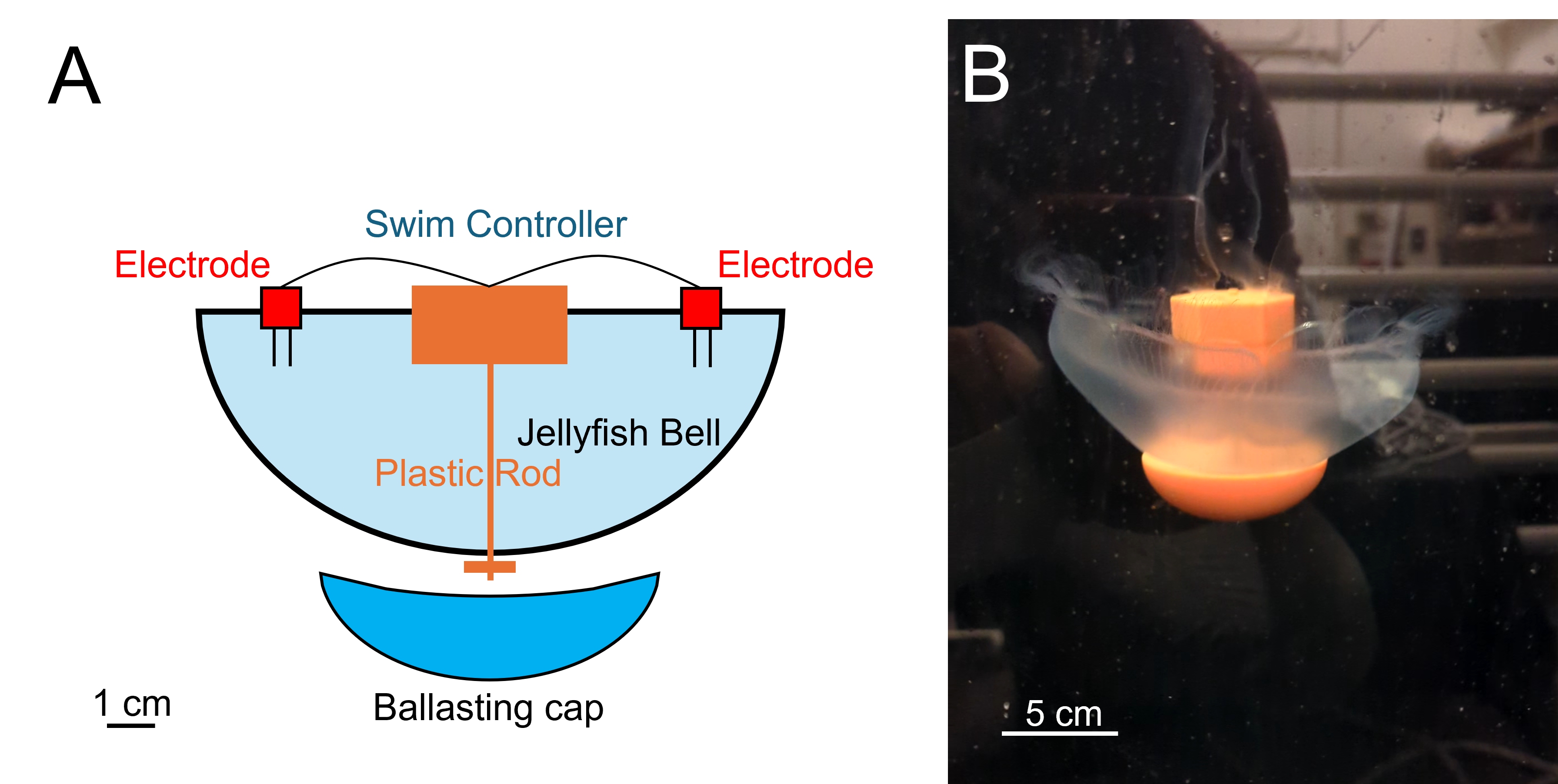}
		\caption{\textbf{Jellyfish stimulation} (A) Schematic of stimulated jellyfish with jellyfish oriented downwards. Consists of two electrodes, a swim controller containing the electronics and battery (not visible), and a plastic rod embedded in the jellyfish tissue with a small threaded nut on the exumbrellar side to keep rod secure. (B) Photo of electronics embedded in live jellyfish during swimming experiment with additional forebody for payload.}
		\label{newschematic}
	\end{figure}
	
	Figure \ref{newschematic}A shows the electronics for onboard electrical stimulation embedded in a model jellyfish. The system consists of a swim controller enclosed in a custom 2.50 ± 0.03 cm diameter electronics housing, two electrodes embedded in the jellyfish muscle tissue, a threaded plastic rod, and a threaded ballasting nut. See previous work for more details on electronics and impact on swimming performance \citep{anuszczykElectromechanicalEnhancementLive2024}. Microscale experiments utilized a small 3D-printed nut on the exumbrellar side of the jellyfish to keep the electronics embedded, while macroscale experiments used a ballasting cap to constrain swimming to vertical. In each case, weights were adjusted to maintain slight positive buoyancy. Figure \ref{newschematic}B shows the electronics stimulating a swimming jellyfish in the lab environment.
	
	\begin{figure*}[!htbp]
		\centering\includegraphics[width=.8\textwidth]{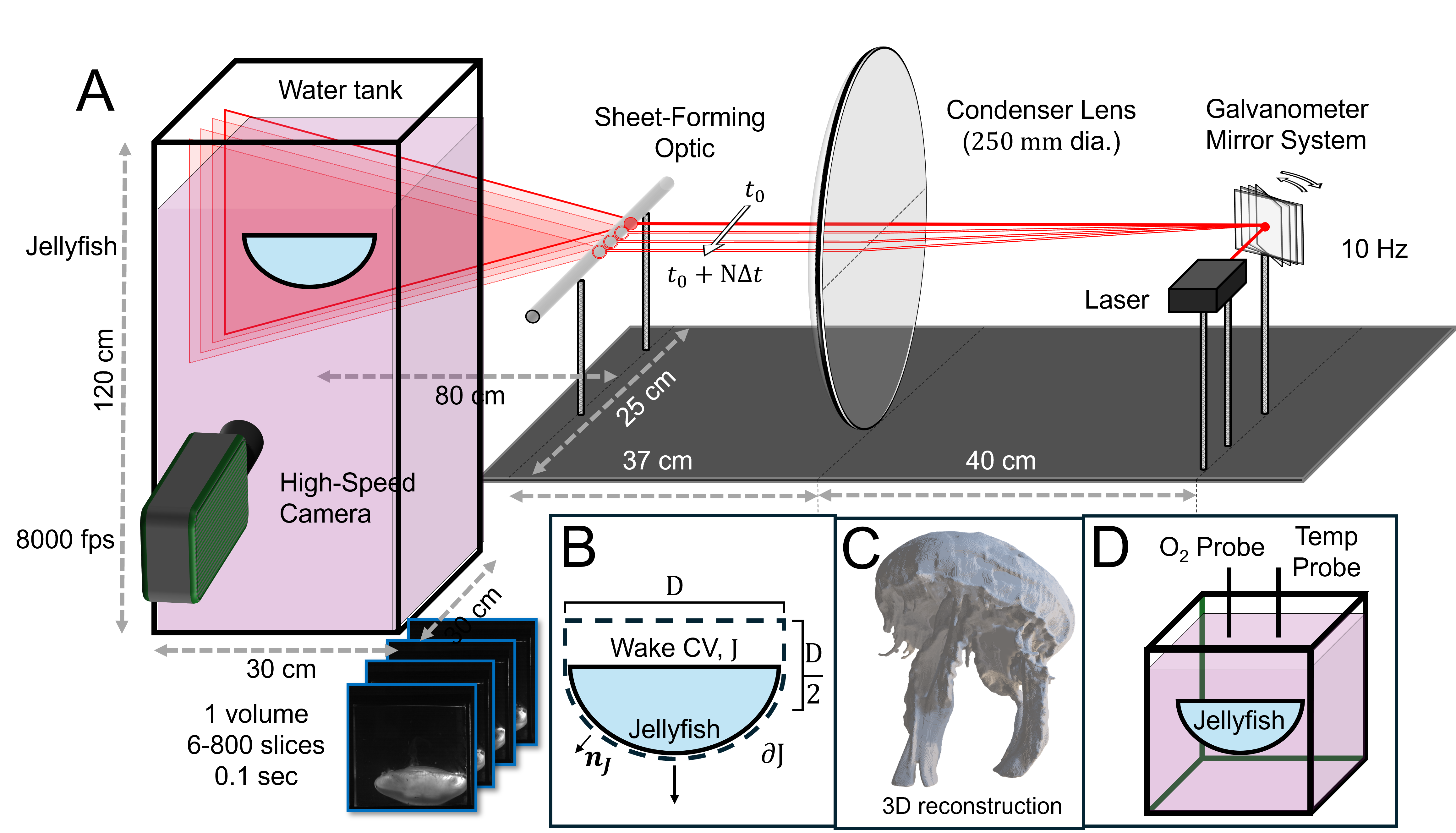}
		\caption{\textbf{Laser scanning system for 3D-3C PIV.} (A) Diagram of laser scanning system with laser, galvanometer rotating at 10 Hz, 250 mm diameter condenser lens to redirect the laser normal to the scanning direction, cylindrical glass rod to form the laser beam into a thin sheet, water tank, and jellyfish. As the laser scans across the jellyfish, a high speed camera records an image stack, or "volume", of 800 slices every 0.1 s. (B) A wake energy volume was defined behind the swimming jellyfish inside which kinetic energy was analyzed. Instantaneous measurements of kinetic energy inside this volume served as a proxy measurement of energy expended into this region of the wake by the swimming animal. The dimensions of the volume were defined as a width of one diameter $D$ and height including the entire animal as well as extending into the posterior wake by $\frac{D}{2}$ from the vertical center of the animal. (C) A representative 3D volumetric reconstruction of a live jellyfish. (D) Closed system respirometry setup used to validate laser scanning technique. Jellyfish oxygen consumption was measured using an O$_2$ probe and temperature probe over time in a sealed chamber. Figure adapted with permission \citep{fu_single-camera_2021}.}
		\label{scanning}
	\end{figure*}
	
	\subsection{Microscale quantitative wake measurements}
	
\subsubsection{Laser scanning PIV}
PIV data of the swimming jellyfish was taken using a three-dimensional, three-component volumetric single-camera laser scanning system, shown in Figure \ref{scanning} (see Supplementary methods for full system details). Volumes were acquired at 10 Hz with sufficient spatial resolution to resolve near wake structures. The system was calibrated using a 3D reference grid to recover spatial scaling.
	
\subsubsection{Post-processing}
Velocity fields were obtained using a custom masking pipeline and processed using OpenPIV (see Supplementary methods for full details).
	
\subsubsection{Analytical analysis}
In order to derive a relationship between the wake motion and the energy consumption of the swimming animal, we first apply the equations of motion to a control volume surrounding the jellyfish. We adapt a model from previous work on vortex propulsion \citep{ruizVortexenhancedPropulsion2011}. Figure \ref{scanning}B illustrates a swimming jellyfish with an associated control volume $J$. The wake energy volume location was defined with a width of one animal diameter $D$ and height including the animal and extending from the vertical center of the animal one radius $\frac{D}{2}$ downstream in the -Y direction. This placement was set to capture the region containing the vortex ring cores during contraction and pinch-off, which scale with bell diameter, as well as flow inside of the bell. Expressing the control volume dimensions in terms of animal diameter $D$ ensures consistent non-dimensional scaling across animals of different morphologies while providing a balance between capturing the energetic wake structures and minimizing inclusion of bow wave and far-field background flow.
	
Let $\mathbf{u}$ denote fluid velocity and $\mathbf{n}$ the outward unit normal to the surface. Assuming incompressibility, the equations of conservation of mass, streamwise (along $y$) momentum, and energy are then given by

\begin{equation}
\begin{aligned}
\text{mass:} 
&\quad 0 = \rho \oint_{\partial J(t)} \mathbf{u} \cdot \mathbf{n}\, dA
\\[6pt]
\text{$y$-momentum:}
&\quad F_{\mathrm{ext},y}
= \rho\left[\frac{d}{dt}\int_{J(t)} v\, dV
+ \oint_{\partial J(t)} v\,\mathbf{u}\cdot\mathbf{n}\, dA\right]
\\
&
\quad\qquad + \oint_{\partial J(t)} p\, n_y\, dA
\\[6pt]
\text{energy:}
&\quad \dot{W}_{\mathrm{ext}}
= {\rho\over 2}\!\left[
{d\over dt}\!\int_{J(t)}\!\|\mathbf{u}\|^2\,dV
+ \!\oint_{\partial J(t)}\!\|\mathbf{u}\|^2\mathbf{u}\!\cdot\!\mathbf{n}\,dA
\right]
\\
&
\quad\qquad + \oint_{\partial J(t)} p\,\mathbf{u}\cdot\mathbf{n}\, dA
\end{aligned}
\label{eq:cv_mass_momentum_energy}
\end{equation}
	
where seawater density $\rho=1024$ kg m\textsuperscript{-3} at 35 PPT and 21°C \citep{dabiri_wake-based_2010,li_volumetric_2023} and $\partial{J}$ is the surface of the control volume. $F_{ext,y}$ is the net force exerted by the swimmer on the fluid, $v$ is the y component of velocity vector $\mathbf{u}$, $p$ is the pressure, and $\dot{W}_{\mathrm{ext}}$ is the external mechanical power imparted by the swimmer to the fluid. Because pressure and viscous stresses on the control surface are not directly resolved, a full thrust estimate is not attempted. Instead, the measurable velocity-based portions of the momentum and energy balances are used. The equations of  streamwise momentum and energy thus simplify to
	
\begin{equation}
\begin{aligned}
\text{$y$-momentum:}\qquad
F_{\mathrm{ext},y}
&= \rho\!\left[
{d\over dt}\!\int_{J(t)}\! v\,dV
+ \!\oint_{\partial J(t)}\! v\,\mathbf{u}\!\cdot\!\mathbf{n}\,dA
\right]
\\[6pt]
\text{energy:}\qquad
\dot{W}_{\mathrm{ext}}
&= {\rho\over 2}\!\left[
{d\over dt}\!\int_{J(t)}\!\|\mathbf{u}\|^2\,dV
\right.
\\
&\qquad\left.
+ \!\oint_{\partial J(t)}\!\|\mathbf{u}\|^2\,\mathbf{u}\!\cdot\!\mathbf{n}\,dA
\right]
\end{aligned}
\label{eq:momn}
\end{equation}
	
We distinguish between three related but distinct quantities: the resolved wake kinetic energy within the finite measurement control volume, the total mechanical power imparted by the swimmer to the fluid, and the organismal metabolic energy consumption. These microscale measurements quantify only the first of these and therefore represent a partial, lower bound estimate of hydrodynamic energy transfer rather than a complete measure of mechanical power or metabolic consumption.

We analyze the convective transport term in order to study how momentum is transported out of and around the wake control volume. We define the net streamwise (y-component) momentum transport
	
	\begin{equation}
		\begin{aligned}
			\qquad
			M_s(t)=\rho \oint_{\partial J(t)} v \mathbf{u}\cdot\mathbf{n} dA
		\end{aligned}
		\label{eq:momn_trans}
	\end{equation}

For a pulse beginning at $t_a$ through $t_b$

\begin{align} \label{momn_pulse}
M_{\text{s, pulse}} =\displaystyle \int_{t_a}^{t_b} M_s(t) dt
\end{align}

This momentum transport term $M_{\text{s, pulse}}$ was evaluated across the posterior surface located downstream of the swimmer. This surface captures the directed wake momentum transport associated with propulsion as vortex rings and jet structures primarily leave the bell through the downstream wake. Lateral control surfaces instead capture refill and entrainment flows, which do not represent directed propulsion momentum exchange. This downstream control surface thus provides a direct measure of the momentum transported in the wake by each pulse. To allow comparisons across animals of different sizes, the pulse-integrated momentum was normalized by animal mass

\begin{align} \label{momn_mass-specific}
I_{\text{s, pulse}} =\displaystyle \frac{M_{\text{s, pulse}}}{m}
\end{align}

For time-resolved comparisons across animals, we also define the instantaneous mass-specific downstream momentum flux

\begin{align} \label{momn_mass-specific_time}
I_{\text{s}}(t) =\displaystyle \frac{M_{\text{s}}(t)}{m}
\end{align}
	
The momentum transport analysis above characterizes how fluid impulse is exported through the downstream wake. However, this quantity alone does not capture the energetic content of the wake. From the energy balance in Eq.~\ref{eq:momn}, wake energetics are governed by both the temporal change in kinetic energy within the control volume and the convective transport of kinetic energy across its boundaries. Both terms contribute to the mechanical energy imparted to the wake. Here, rather than evaluating these terms separately, instantaneous kinetic energy contained within the wake control volume is used as a proxy for the resolved portion of wake energetic output within the measurement domain \citep{dabiri_wake-based_2010,li_volumetric_2023,ruizVortexenhancedPropulsion2011,katijaViscosityenhancedMechanismBiogenic2009}. This approximation is justified because the control volume is small relative to the evolving wake, such that wake structures pass through the measurement region over short timescales and kinetic energy does not accumulate over long durations. Under these conditions, the kinetic energy contained within the control volume scales with the combined effects of local kinetic-energy growth and downstream kinetic-energy transport. The local kinetic energy density of the velocity field is defined as
	
	\begin{align} \label{KE}
		e_{\text{kinetic}}=\frac{1}{2} \rho \|\mathbf{u(x,y,z,t)}\| ^{2}
	\end{align}
	
	and the instantaneous kinetic energy contained within the wake control volume is obtained by integrating this quantity over $J$ leading to
	
	\begin{align} \label{Ecv}
		E_{\text{J}}(t)=\displaystyle \int_{{J}} e_{\text{kinetic}}(x, y, z,t) \, dV
	\end{align}
	
	To characterize the cumulative wake kinetic energy over an individual pulse, the wake energy is integrated over the pulse duration
	
	\begin{align} \label{Epulse}
		E_{\text{J, pulse}} =\displaystyle \int_{t_a}^{t_b} E_{\text{J}}(t) dt
	\end{align}
	
	To compare hydrodynamic energy loss across animals and swimming conditions, wake kinetic energy was normalized by the kinetic energy of the swimming jellyfish. This normalization yields a dimensionless measure of how strongly pulse-associated motion is exerted into the surrounding fluid relative to the motion of the animal. Because absolute wake kinetic energy increases with the overall magnitude of swimming motion, the ratio provides a more comparable metric across pulses and individuals than energy alone. These ratios should therefore be interpreted as comparative measures of resolved fluid-energy loss relative to animal motion, rather than as direct estimates of propulsive efficiency or total metabolic expenditure. We define an instantaneous energy ratio 
	
	\begin{align} \label{Eratio}
		E_{\text{ratio}}(t) = \frac{E_J(t)}{E_{\text{jelly}}(t)}
	\end{align}
	
	where $E_{\text{jelly}}(t)$ is the kinetic energy of the animal. The mass of each animal was found by weighing them after conducting experiments. Each animal was carefully removed from the tank, patted dry to remove excess water, and weighed on an electric balance (Eosphorus Sf-400C, 600 g range, 0.01 g precision). The three-component velocity of the swimming animal was found by tracking the movement of the bell apex. The center XY planes bisecting each animal apex were manually identified in PFV4 (Photron, San Diego, CA), and videos were exported to Matlab for analysis. The point tracking tool Dltdv8 \citep{hedrick_software_2008} in Matlab was used to track XY apex locations over time. MATLAB was used to generate full image stack slices along the YZ planes again bisecting the animal bell apex. These were similarly tracked to find Z trajectories. 
	
	Each experimental trial captured several pulses, with stimulated trials including more pulses than unstimulated trials due to higher pulse rates. The analysis was thus manually split into individual pulses by tracking bell margin movement and identifying pulse start and end times $t_a$ and $t_b$. The pulse-specific normalized wake energy ratio was defined as 
	
	\begin{align} \label{Epulse}
		E_{\text{ratio, pulse}} =\frac{E_{\text{J,pulse}}}
		{\displaystyle 
			\int_{t_a}^{t_b} E_{\text{jelly}}(t)\, dt}
	\end{align}
	
	Electrical stimulation is known to increase pulse rates significantly compared to unstimulated jellyfish. While the electronics were used to stimulate at a constant frequency of $f_{ST} = 0.5$ Hz, unstimulated \textit{Aurelia} of similar size are known to pulse at $f_{US} = 0.16$ Hz \citep{xu_low-power_2020}. This difference was corrected for by defining a time-specific normalized wake energy ratio to estimate the rate of energy deposition, or power, expended into the surrounding fluid
	
	\begin{align} \label{Etime}
		P_{\text{ratio, time}} = f E_{\text{ratio, pulse}}
	\end{align}

This metric represents the expected energy consumption of the stimulated and unstimulated jellyfish. 

We follow previous work in estimating the role of kinetic energy dissipation in these measurements \citep{dabiri_wake-based_2010}. Wake vortices are known to break down on timescales of $\frac{d}{U}$ with $d$ the vortex size and $u$ vortex velocity \citep{saffmanVortexDynamics1993}. At the low Reynolds numbers observed here, this process is negligible and the time scale of viscous dissipation can be found through dimensional analysis \citep{rosenheadLaminarBoundaryLayers1963} of 
	
	\begin{align} \label{diss}
		\tau = \frac{d^2}{\nu}
	\end{align}
	
These experiments have eddy sizes larger than 12 cm and kinematic viscosity of $10^{-2}$ cm$^2$ s$^{-1}$ \citep{batchelorIntroductionFluidDynamics2000} leading to a dissipation timescale of $\tau$ much larger than typical pulse durations on the order of 1 second. 
	
\subsubsection{Biomechanics analysis}
We conducted experiments focused on understanding the impact of electronic stimulation on the biomechanics of jellyfish. These experiments investigated both the impact of stimulation as well as the embedded electronics without stimulation. Margin movement was measured with four different treatments. The treatments were: the unstimulated animal, the unstimulated animal with passive electronics embedded, the stimulated animal, and a final repeated unstimulated animal experiment. The additional unstimulated treatment allowed the measurement of how animals might return to baseline biomechanics after stimulation experiments. Margin movement experiments were conducted with three additional animals, with one 10 second trial analyzed for each of the four treatments. While additional trials were recorded, only trials where the animal remained in the scanning volume throughout the duration of the experiment were analyzed.
	
To interpret changes in wake energetics, we quantified jellyfish biomechanics under the same experimental conditions. The biomechanics of the swimming animals were analyzed in two dimensions at the center plane of the animal. DLTdv8 was used to track both the animal apex and margins, enabling the calculation of biomechanics metrics across treatments. Margin movement was defined as the distance traveled by a single margin between a maximally open or relaxed bell state and maximally closed or contracted bell state. This was normalized by body length, defined as the projected maximum diameter of each animal. Contraction and relaxation duration and speed were measured from these margin-tracking trials as well as the volumetric PIV experiments, for a total of six animals. Contraction speed was additionally normalized by body length. These metrics are relevant both for measuring energy expenditure and as guides for informing optimization of electrical stimulation strategies.

	\subsubsection{Statistical Analysis}
	For each trial, pulse-level metrics were first calculated over manually identified pulse windows in MATLAB. Analyses were restricted to complete pulses, defined from maximum relaxation through contraction to the subsequent maximum relaxation; partial pulses were excluded prior to averaging. For each trial within each animal, the pulse-level values were averaged to obtain  a single trial-level mean for that metric. Statistical comparisons between unstimulated and stimulated conditions were performed on these trial means using paired $t$-tests, with each unstimulated trial paired to the corresponding stimulated trial from the same animal. Thus, the unit of replication for hypothesis testing was the animal-trial pair, not the individual pulse. Statistical comparisons between unstimulated and stimulated conditions for contraction duration and speed were performed on trial-averaged values using Welch’s two-sample $t$-test to account for unequal sample sizes and variances. The first and last unstimulated treatments exhibited minimal differences and were therefore pooled into a single unstimulated group for analysis. Pulse-level points shown in Figures were included to illustrate intra-animal variability, but statistics were based on trial-level means. Statistical significance was defined as $p<0.05$.
	
	\subsection{Macroscale free-swimming energy measurements}
	
\subsubsection{Laser scanning for volumetric reconstruction}
Volumetric reconstructions of the jellyfish were created using the same single-camera laser scanning system as shown above in Figure \ref{scanning} (see Supplementary methods for details).
	
\subsubsection{Validation of volumetric reconstruction}
Experiments were performed to characterize the accuracy and precision of the laser scanning technique. To measure the relationship between volumetric reconstructions and wet weight of jellyfish, 10 jellyfish were scanned and weighed to compare wet weight (WW) and laser scanned weight (LSW). Each jellyfish was first scanned using the procedure described in this section. Then, they were carefully removed from the tank, patted dry to remove additional water, and weighed on an electric balance (see Supplementary methods for details).
	
\subsubsection{Energetic consumption from volumetric changes}
This laser scanning technique used repeated laser scans and volumetric reconstructions over several hours to convert a change in volume due to catabolysis to an average energy consumption rate for each jellyfish. This method was validated against simultaneous respirometry and agreement was found (see Supplementary methods).

The method to convert the measured change in laser-reconstructed animal size to an energetic rate was based on earlier animal physiology work \citep{harrisICESZooplanktonMethodology2000}. The respiratory rate is defined as
	
	\begin{align}\label{RR_ikeda}
		RR=\frac{\Delta V\rho C}{RQa}
	\end{align}
	
in terms of the change in volume of the animal $\Delta V$, tissue density $\rho$, tissue carbon content $C$, respiratory quotient $RQ$, and a unit conversion term $a$. The change in volume $\Delta V$ was measured using the laser scanning method described in this section. Since jellyfish are neutrally buoyant and approximately 96\% water \citep{uyePopulationBiomassFeeding2005}, $\rho$ was assumed to be equal to the density of seawater $\rho=1024$ kg m\textsuperscript{-3} at 35 ppt and 21°C. We experimentally measured the tissue carbon content C of 9 \textit{Aurelia aurita} (see Supplemtary methods). The respiratory quotient $RQ$ depends on the type of tissue metabolized and represents the ratio of carbon dioxide produced to oxygen consumed. For jellyfish, we used a range $RQ=0.8-0.9$ since they metabolize primarily protein \citep{uyePopulationBiomassFeeding2005,harrisICESZooplanktonMethodology2000}. Uncertainty in $RQ$ (0.8–0.9), tissue carbon content, and volumetric reconstruction was propagated to estimate bounds on metabolic rate. These uncertainties primarily affect absolute metabolic estimates but have a reduced impact on relative comparisons between experimental conditions. Previous work reports that the weight specific respiratory rate of \textit{Aurelia aurita} does not change with starvation state, \citep{frandsenSizeDependentRespiration1997} suggesting Equation \ref{RR_ikeda} does not need to be corrected for starvation state. We normalize all $RR$ by animal $WW$ in grams to allow comparisons across animal morphology.
	
	\subsubsection{Electrical stimulation for continuous swimming}
	During long-duration free-swimming experiments, a ballasting cap was added to passively maintain vertical downward swimming by lowering the center of gravity of the animal. This additional volume could house future ocean sensors. See previous work \citep{anuszczykElectromechanicalEnhancementLive2024} for more details on electronics and ballasting cap design and impact on swimming performance. Figure \ref{newschematic}B shows the electronics stimulating a swimming jellyfish in the lab environment. The simultaneous laser scanning and respirometry experiments described in this section were additionally performed with 3 electrically stimulated animals. 
	
\subsubsection{Macroscale stimulated free-swimming experiments}
	
	\begin{figure*}[!h]
		\centering\includegraphics[width=.8\textwidth]{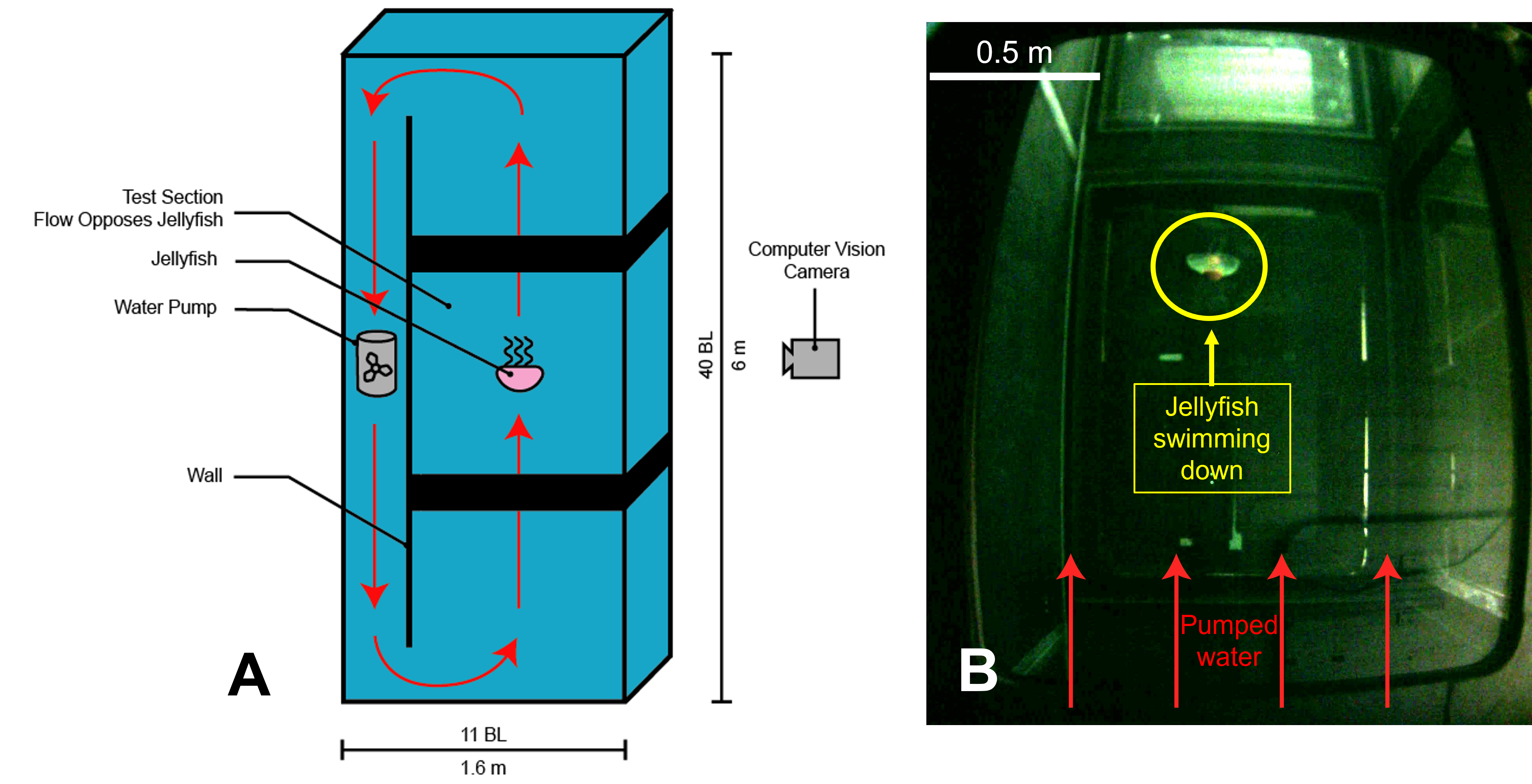}
		\caption{\textbf{Free-swimming tank for long-duration experiments.} (A) Diagram of 6 m by 1.6 m artificial seawater vertical tank facility. Water pumps in the recirculation area on the left generated flow (shown with red arrows), which opposed free-swimming jellyfish. A computer vision camera tracked the swimming animal. (B) The view from the computer vision camera of the middle section of the tank as a jellyfish, circled in yellow, swam downward. As the jellyfish descends, the camera ran a PID control loop interfacing with the water pumps to control the flow, keeping the animal in view of the camera.}
		\label{vert}
	\end{figure*}

Free-swimming experiments were conducted in a 6-m tall vertical tank with flow control enabling continuous swimming (Figure \ref{vert}A). Animal position was tracked in real time to maintain station, and swimming speed was determined from combined body motion and flow velocity (see Supplementary methods). Figure \ref{vert}B shows the view from the tracking camera of the middle section of the tank as a jellyfish was swimming downwards.
	
\subsubsection{Basal metabolism}
Basal metabolism was measured using immobilized animals (see Supplementary methods), allowing separation of basal and swimming contributions.
	
\subsubsection{Swimming metabolism model}
We derived a physics-based quasi-steady swimming metabolism model based on previous work \citep{dabiriFastswimmingHydromedusaeExploit2006,acuna_faking_2011} to provide an order-of-magnitude framework for interpreting dominant physical contributions to measured energy consumption. Here, we model the jellyfish energy consumption $\frac{dE}{dt}$ as the sum of the basal metabolism $P_{basal}$, the power expended setting the surrounding water into motion while swimming $P_{wake}$, and the hydrodynamic drag on the free-swimming animal $P_{drag}$
	
	\begin{align} \label{energybalance}
		\frac{dE}{dt}=P_{basal}+P_{wake}+P_{drag}
	\end{align}
	
	We model the water motion created by the animals while swimming as primarily comprising a vortex ring. Following previous work \citep{dabiriFastswimmingHydromedusaeExploit2006}, we approximate its kinetic energy as that of an equivalent oblate spheroid of water set into motion each time the jellyfish pulses and generates a vortex ring. We use the classical kinetic energy formula for the kinetic energy contained in the wake,
	
	\begin{align}
		KE_{wake}=\frac{1}{2}MU^2_{wake}
	\end{align}
	
	where $M$ is the mass of water set in motion and $U_{wake}$ is the speed at which this wake convects. The mass is given by $M=\rho V$ where seawater density $\rho=1024$ kg m\textsuperscript{-3} at 35 PPT and 21°C, and the volume $V=\frac{4}{3}\pi R^2h$ where R is the radius of the animal and h is the semi-minor axis of the spheroid. As the vortex ring is accelerated, it accelerates the surrounding fluid, necessitating an added mass term $C_{am}$. We also introduce an effective swimming efficiency term $\eta$ to account for energy lost in physiological processes that convert body motion to fluid motion leading to
	
	\begin{align}
		KE_{wake}=\frac{2}{3\eta}\rho \pi R^2h(1+C_{am})U^2_{wake}
	\end{align}
	
	Power is the time rate of change of kinetic energy. \textit{Aurelia aurita} generate two vortex rings with each swimming cycle \citep{gemmell_passive_2013}. Thus, we assume that the kinetic energy for two vortex rings is expended with each swimming cycle. Hence,
	
	\begin{align} \label{pwake}
		P_{wake}=\frac{4}{3\eta}\rho \pi R^2h(1+C_{am})U^2_{wake}f_{pulse}
	\end{align}
	
	where $f_{pulse}$ is the pulse frequency. We model the steady drag on the free-swimming jellyfish as the product of the drag force and the swimming speed divided by effective swimming efficiency. The unsteady effects are captured with the addition of a potential flow term \citep{lighthillFundamentalsConcerningWave1986}
	
	\begin{align} \label{pdrag}
		P_{drag}=\frac{1}{2\eta}C_D\rho SU^3_{swim} + maU_{swim}C_{am}
	\end{align}
	
	with $C_D$ the drag coefficient for the jellyfish, $S$ the projected planform area in the direction of swimming, $U_{swim}$ the swimming speed of the animal, $m$ the mass of the jellyfish, and $a$ the acceleration of the animal across a swimming cycle. Thus, combining Equations \ref{pwake} and \ref{pdrag} into \ref{energybalance} becomes 
	
	\begin{equation}\label{dedt}
		\begin{split}
			\frac{dE}{dt} = P_{basal} + \frac{\rho}{\eta} \Big[
			\frac{4}{3}\pi R^2 h (1+C_{am}) U^2_{wake} f_{pulse} \\
			+ \frac{1}{2} C_D S U^3_{swim}
			\Big] + ma U_{swim} C_{am}
		\end{split}
	\end{equation}
	
	Jellyfish geometry was measured to determine $R$, and $h$ was estimated as $h=0.4R$. The stimulated rate was set by the electronics at $f_{pulse, stim}=0.5$ Hz, and previous work found jellyfish natural pulse rates of $f_{pulse, nat}=0.16$ Hz \citep{xu_low-power_2020}. While previous studies report hydrodynamic efficiencies on the order of $0.2-0.4$ \citep{acuna_faking_2011}, these values do not account for internal energy losses within the animal, including viscoelastic dissipation and muscle inefficiency. Accordingly, we use a range of $\eta=0.1-0.3$ as a representative effective swimming efficiency. $C_{am}$ was assumed to be 1, and for a flat plate $C_D=1.12$ \citep{hoernerFluidDynamicDrag}. The wake convection speed $U_{wake}$ was approximated as the contraction velocity of the bell margin, measured by analyzing a naturally pulsing animal, and $U_{swim}$ for the free-swimming experiments was measured as described in this section. To compare with respirometry data, we assumed 447 kilojoules per mole of oxygen consumed for protein, 12/22.4 is the weight of carbon in 1 mole of carbon dioxide \citep{harrisICESZooplanktonMethodology2000}, and there are $2.4\times10^4$ ml per mole of oxygen \citep{garciaOxygenSolubilitySeawater1992}. To normalize by wet weight mass, the relation
	
	\begin{align} \label{ww_radius}
		WW=0.547R^{2.86}
	\end{align}
	
	was used \citep{uyePopulationBiomassFeeding2005}.

This model is not intended to provide predictive accuracy, but rather to contextualize experimental observations and estimate the relative contributions of wake generation, drag, and basal metabolism.

\section{Results}
	
\subsection{Microscale quantitative wake measurements}
\subsubsection{Laser scanning PIV}
We measured the 3D-3C flow induced by swimming jellyfish using the laser scanning system described above. The dominant flow structures occur in the trailing wake, therefore only the downstream wake region is analyzed here.
	
	\begin{figure}[!htbp]
		\includegraphics[width=0.5\textwidth]{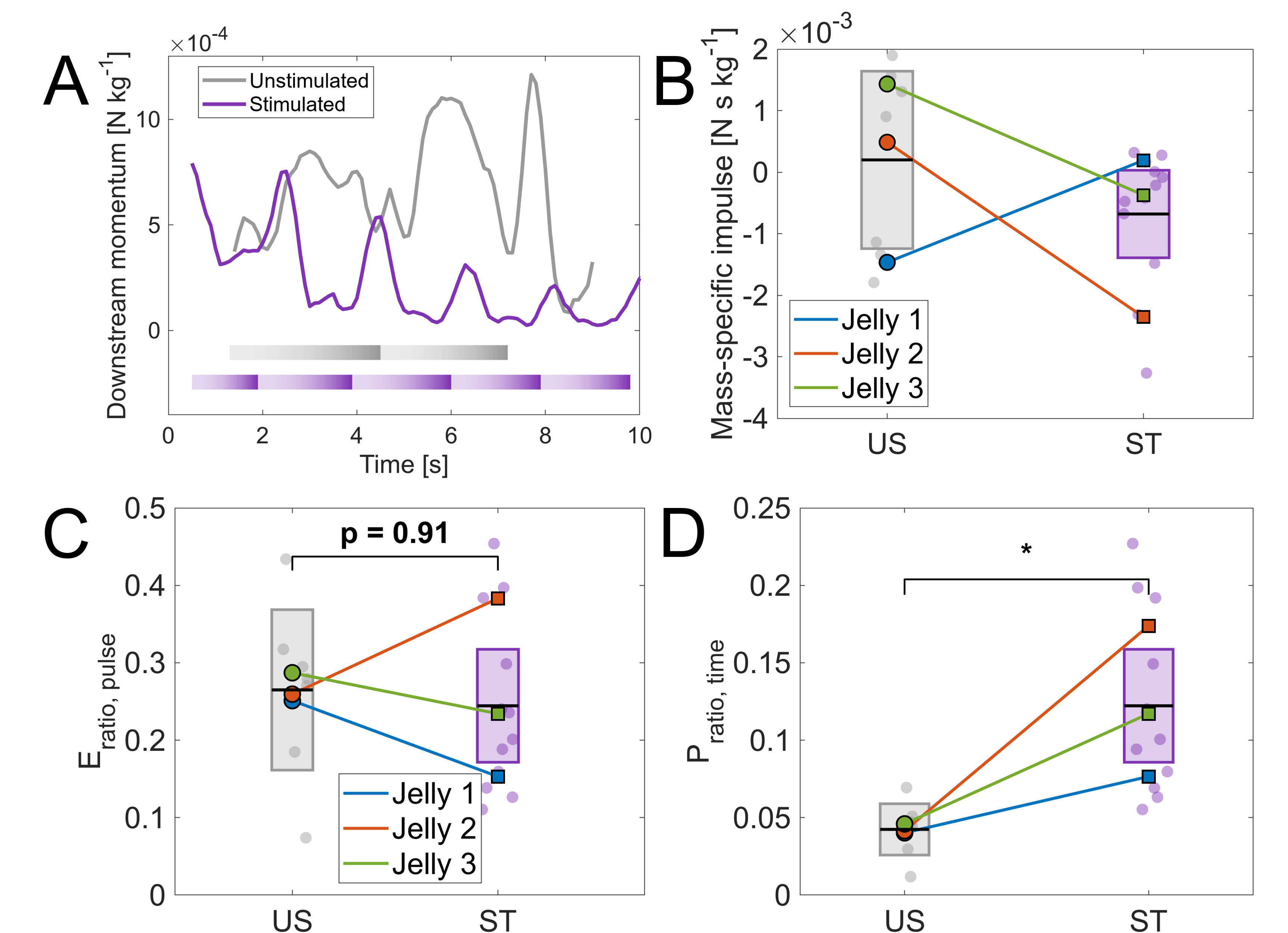}
		\caption{\textbf{Normalized wake energy.} (A) Instantaneous mass-specific net downstream momentum, Equation \ref{momn_mass-specific_time}, transported during each pulse for an unstimulated (gray) and stimulated (purple) animal over the course of 10 second experiments. This trace shows two unstimulated pulses and five stimulated pulses indicated at the bottom of the plot by the colored gradient from light to dark where light represents the beginning of the contraction. (B) Mass-specific net downstream momentum, equation \ref{momn_mass-specific}, transported during each pulse. Points represent individual pulses and colored markers denote animal means for unstimulated (US) and stimulated (ST) trials connected by lines for each animal. Shaded boxes indicate the 95\% confidence interval of the pooled pulse mean for each condition. (C) Pulse-specific normalized energy $E_{\text{ratio, pulse}}$ given by eq. \ref{Epulse}. Blue, red, and green points represent animal means for each condition and shaded boxes denote the 95\% confidence interval of the pooled pulse mean for each condition. Gray and purple points represent individual pulses. Statistical comparisons were performed on trial-level means using paired $t$-tests, pairing US and ST trials from the same animals. No significant difference in energy released per pulse was observed between conditions ($p=0.91$). (D) Time-specific normalized energy $P_{\text{ratio, time}}$ defined by Equation \ref{Etime}. Stimulated jellyfish release 2.9 $\pm$1.2 times more energy per second to the near wake than unstimulated animals ($p=0.047$). Points and shaded boxes are as in panel B.}
		\label{energy}
	\end{figure}
	
	Figure \ref{energy}A compares an unstimulated and stimulated trial of the same jellyfish. Here we show the instantaneous mass-specific net downstream momentum within the wake energy control volume at each instant in time over the course of the 10 second experiments. The unstimulated animal downstream momentum, shown here in gray, shows 2 peaks with one appearing at each pulse. The color gradient bars at the bottom of the Figure show each pulse going from light to dark, where light represents the beginning of the contraction. The stimulated animal shows 5 pulses over the same period of time due to the consistent pulse frequency of the electronics.

Pulse-specific normalized energy, Figure \ref{energy}B, was not significantly different between unstimulated and stimulated conditions (p = 0.91), indicating that the energy imparted per pulse remains approximately constant. However, stimulated animals exhibited a 2.9 ± 1.2 increase in time-normalized energy due to their higher pulse frequency in Figure \ref{energy}C. These results indicate that increased energetic output under stimulation is driven primarily by increased pulse rate rather than changes in per-pulse hydrodynamic efficiency.
	
	\subsubsection{Biomechanics Analysis}
	
	\begin{figure}[!tbp]
		\centering\includegraphics[width=.5\textwidth]{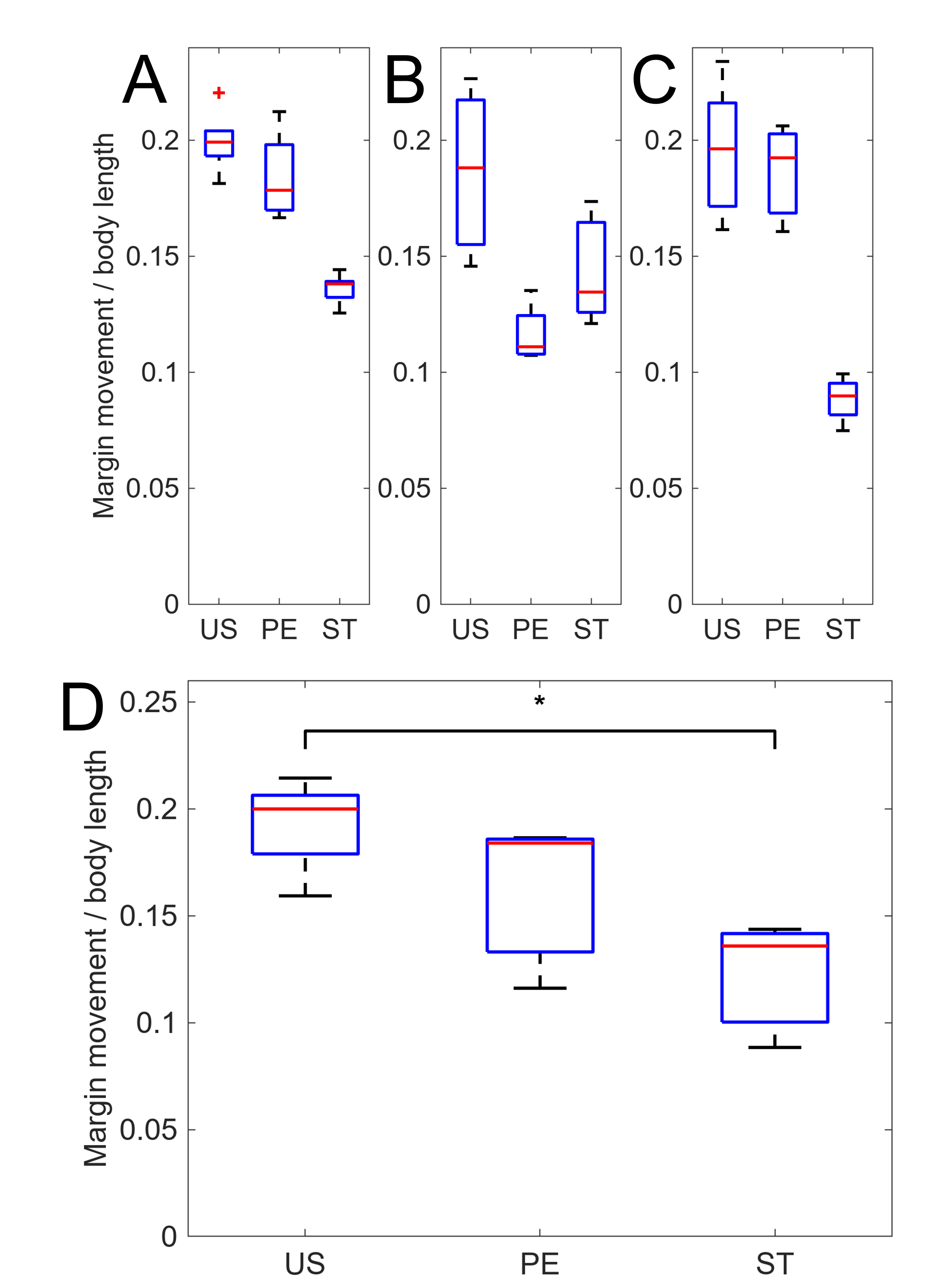}
		\caption{\textbf{Margin movement} (A)-(C) Margin movement between fully relaxed and contracted states for three jellyfish with diameters of 15$\pm$0.1 cm, 19$\pm$0.1 cm, and 27$\pm$0.1 cm respectively, measured at full relaxation. "US", "PE", and "ST" refer to unstimulated, unstimulated with embedded onboard passive electronics, and stimulated experiments. Stimulation reduces margin movement with a p-value of 0.035, while there is no statistically significant difference between the other experiments. (D) Average margin movement across (A)-(C). Stimulation reduces bell margin movement by 36.5$\pm$5.4\% with a p-value of 0.035.}
		\label{margin}
	\end{figure}
	
	We compare swimming animals across different treatments and observe that electrical stimulation changes swimming biomechanics. Figures \ref{margin}A-C show the margin movement between maximum relaxation and contraction states normalized by the body diameter for three different jellyfish. These animals span a wide range of adult medusae sizes with Figures \ref{margin}A-C having diameters measured at full relaxation of 15$\pm$0.5 cm, 19$\pm$0.5 cm, and 27$\pm$0.5 cm, respectively. As above, "US" represents the unstimulated trials and "ST" is the stimulated trials, while "PE" is measurements conducted with passive electronics embedded in the animal but without electrical stimulation. No change was observed with a final unstimulated animal experiment, so these results were combined with the initial unstimulated measurements. The "US" trials involved measuring an average of 5.7$\pm$0.5 complete pulses defined as full relaxation to full contraction, the "PE" trials involved measuring an average of 5.7$\pm$0.5 pulses, and the "ST" trials involved measuring an average of 5.3$\pm$1.2 pulses. 
	
Figure \ref{margin}D averages across the three animals shown in Figures \ref{margin}A-C and shows a statistically significant decrease in margin movement with stimulation. The average of the margin movement from Figures \ref{margin}A-C for each treatment is shown with the "US" and "ST" treatments resulting in a p-value of 0.035 measured using Welch's $t$-test. Comparing "US" and "PE" was not significant with a p-value of 0.31, while "PE" and "ST" have a p-value of 0.25. This shows that these embedded electronics stimulating the swimming jellyfish limit bell margin movement. 
	
We also measured contraction and relaxation duration and speed for six jellyfish comparing unstimulated and stimulated experiments. The relaxation duration exhibited a statistically significant decrease with a p-value of 0.042, while contraction duration was not statistically significant with a p-value of 0.17. Stimulation decreases relaxation duration by 24$\pm$9.7\%. These measurements agree with previous observations of unstimulated jellyfish contraction and relaxation durations. In addition, animal biomechanics would predict this result, since jellyfish bell contraction is an active muscle movement while relaxation is mostly passive \citep{gemmell_passive_2013,costello_hydrodynamics_2021}. Neither contraction nor relaxation speeds changed with stimulation, suggesting that the muscular response from electrical stimulation is not different from the muscular response in an unstimulated pulse. We found p-values for contraction and relaxation speeds of 0.73 and 0.9, respectively.
	
\subsubsection{Laser scanning technique validation}
The volumetric reconstruction method was validated against wet weight measurements and respirometry, showing no significant differences between methods (see Supplementary Figure 10).
	
\subsubsection{Long duration stimulated free-swimming experiments}

\begin{figure}[!h]
	\centering\includegraphics[width=.45\textwidth]{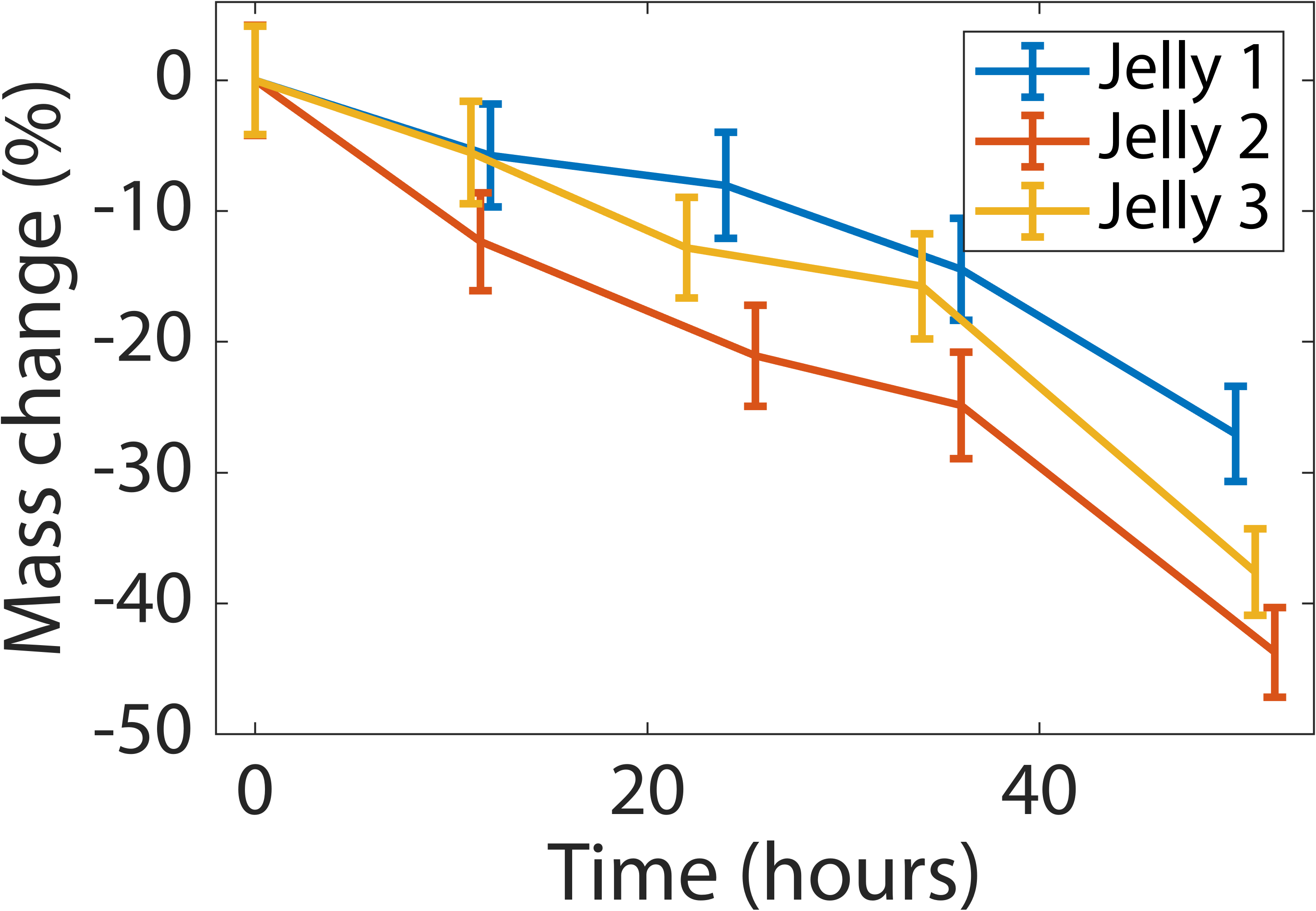}
	\caption{\textbf{Observed mass decrease due to catabolysis.} We deployed this technique to measure free-swimming metabolism during 50 hour experiments for 3 jellyfish. The jellyfish lost 36\% ± 8.4\% of their mass on average over the course of this experiment. All error bars represent 1 standard deviation.}
	\label{mass}
\end{figure}

We utilized this laser scanning technique to measure the free-swimming metabolism of 3 electrically stimulated, continuously swimming jellyfish in a 6-m tall vertical tank. Figure \ref{mass} shows the mass change of each animal as a percent of the mass at the previous timestep. The error was estimated by taking repeated scans of the animal. The continuously swimming animals lost a mean of 36\% ± 8.4\% of their mass over the course of 50 hours, which is in line with previous results on starvation \citep{ishiiFoodRegulationGrowth1998}. The starting masses of the animals found from scanning were $m_{jelly, 1}=470 \pm 19$ g, $m_{jelly, 2}=347\pm14$ g, and $m_{jelly, 3}=536\pm21$ g. We found that the animals swam a mean distance of 2.55 ± 0.46 km while in view of the computer vision camera at speeds of 2.44 ± 0.60 cm s$^{-1}$. Previous results showed similar electrically stimulated swimming speeds over much smaller timescales on the order of minutes \citep{anuszczykElectromechanicalEnhancementLive2024}.
	
	\begin{figure}[!h]
		\centering\includegraphics[width=.5\textwidth]{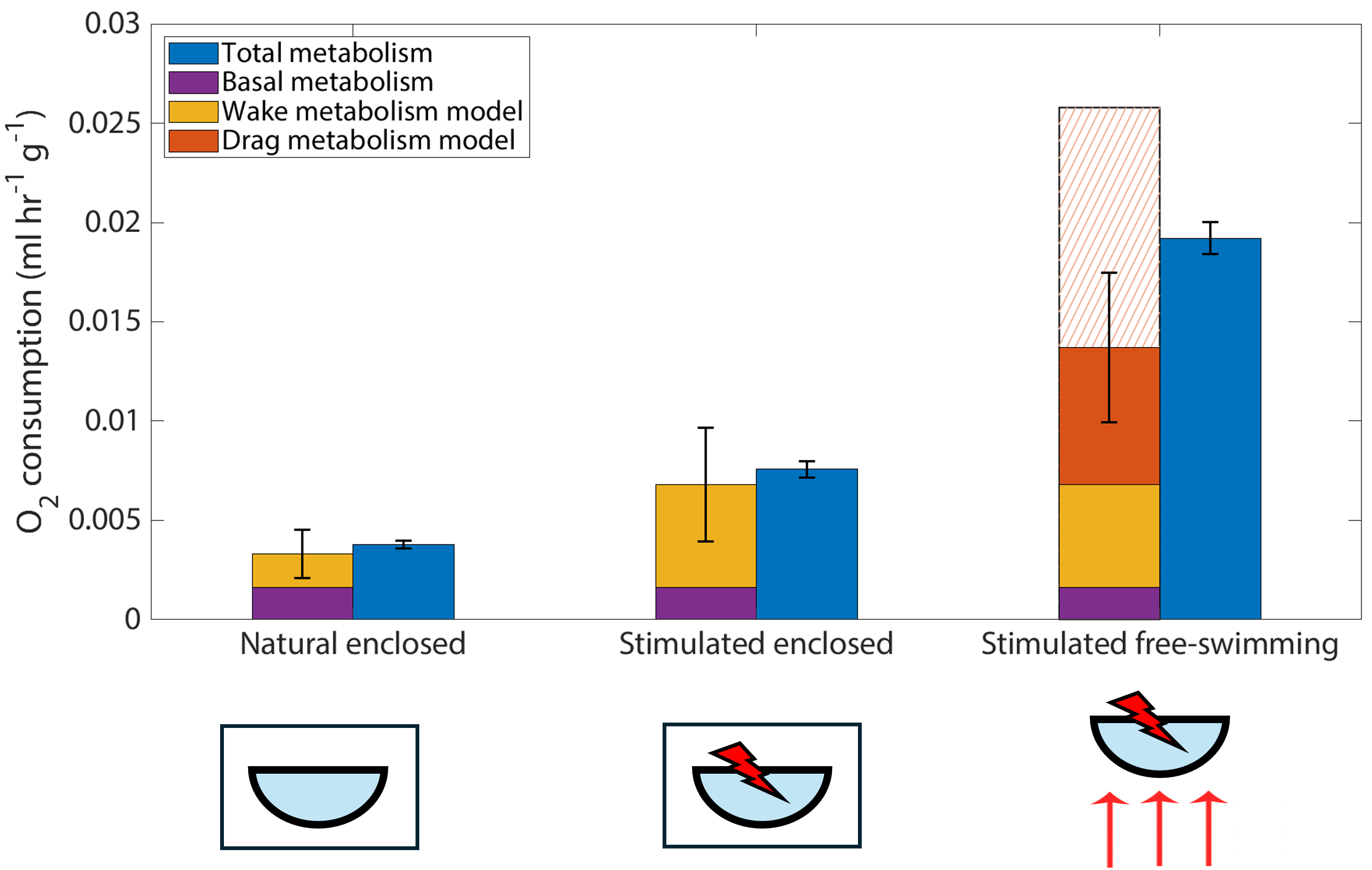}
		\caption{\textbf{Free-swimming energy consumption and metabolism model.} Stimulated free-swimming jellyfish consume 2.5 times as much energy as stimulated enclosed animals. The wake metabolism model represents the energy to set wake into motion. Hatched bar in drag term represents the maximum consumption where potential term in Equation \ref{pdrag} uses maximum acceleration measured.}
		\label{vert_energy}
	\end{figure}
	
Free-swimming experiments showed increased energy consumption. Figure \ref{vert_energy} compares the energetic costs of swimming jellyfish in the three different configurations tested. The blue bars represent the total metabolism. In the “Natural enclosed” configuration and the “Stimulated enclosed” configuration, this was measured in the respirometry chamber with the oxygen probe, and in the “Stimulated free-swimming” configuration using the laser scanning technique in the vertical tank. We found that the free-swimming animal consumes 2.5 times more energy, normalized by body mass, than the stimulated animal in the enclosed respirometry chamber. This increase exceeds the combined uncertainty bounds associated with the volumetric method, indicating that the observed difference reflects a robust change in energetic cost rather than measurement variability. This increase reflects multiple differences between configurations, including swimming speed, flow environment, and behavioral state. In particular, free-swimming animals move at finite speed, experience reduced recirculation effects, and actively maintain position, all of which contribute to increased energetic cost. While validation was performed on a limited number of animals (n = 3), agreement with respirometry measurements supports the use of this method for comparative analysis. Because hydrodynamic power scales as the cube of swimming speed \citep{Daniel1983MechanicsEnergeticsMedusanb}, small increases in speed result in disproportionately higher energetic costs. The increase in swimming speed from nearly zero in the enclosed configurations to 2.44 ± 0.60 cm s$^{-1}$ in the free-swimming configuration would thus increase the hydrodynamic power expended. In addition, recirculation in a small respirometry chamber reduces the force required to contract the bell, suggesting that the same movement might require more energy during free-swimming. Because these factors vary simultaneously, the observed increase cannot be attributed solely to hydrodynamic drag. 
	
We found an average basal metabolism of $P_{basal} = (1.6 \pm 0.4)\times10^{-3}$ ml O$_2$ hr\textsuperscript{-1} g\textsuperscript{-1} which constitutes 42\% ± 10\% of jellyfish natural swimming energy consumption based on our experiments with rhopalia excision. This is shown in Figure \ref{vert_energy} in purple for each test configuration. This aligns with previous work for \textit{Stomolophus meleagris} which found a basal metabolic rate of 50\% \citep{larson_respiration_1987}. Jellyfish are known to be among the most efficient metazoans in terms of cost of transport (COT), defined as the mass-specific energy per distance traveled \citep{gemmell_passive_2013}. Their low basal metabolism compared to other species suggests that their efficiency is not only due to efficient swimming as previously proposed. Instead, the low COT is also due to highly efficient essential physiological functions that make up basal metabolism, such as digestion and reproduction.
	
Our swimming metabolism model is shown in Figure \ref{vert_energy} in orange alongside each configuration with error bars representing a range of effective swimming efficiencies. The drag term is only shown for the free-swimming animal in red, as described in Methods. This model generally captures the order-of-magnitude trends across configurations with the potential term from Equation \ref{pdrag} shown varying based on the acceleration estimate used.

Comparing the resolved wake energy rate with measured metabolic power indicates that the wake measurements capture only a fraction of total energy expenditure. Order-of-magnitude estimates suggest that the resolved wake contribution is substantially smaller than total metabolic power, consistent with unmeasured contributions from pressure work, viscous dissipation, and internal physiological losses.
	
\section{Discussion}
In this paper, we present power measurements of free-swimming jellyfish on both microscale and macroscale timescales. On the microscale, we used three-dimensional PIV to measure full velocity fields within the near-wake region behind free-swimming jellyfish---thus enabling the calculation of momentum and kinetic energy released to this region. We found that electrical stimulation significantly increased the power in the wake on a time-averaged basis by a factor of 2.9$\pm$1.2. On the macroscale, we used a custom, 6-m tall water treadmill and non-invasive laser scanning system to measure total metabolism. We found electrically stimulated free-swimming jellyfish total metabolism was 2.5 times higher than similarly stimulated animals in a closed respirometry chamber. 
	
	Tracking swimming kinematics evinced reduced radial motion with stimulation and a decrease in relaxation duration. This suggests a combination of both electrical and mechanical factors affecting bell contraction. After the initial electrical bell stimulation, the jellyfish contraction speed and duration do not change. After the conclusion of the contraction phase, the relaxation begins and continues at a normal speed but does not reach full relaxation. This reduction in relaxation may arise either from the timing of electrical stimulation preventing full bell relaxation, or from mechanical effects of the embedded electrode wires on the bell margin. Electrical stimulation likely changes the timing of contraction relative to the development of the starting vortex, which may reduce the efficiency with which momentum is transferred to the surrounding fluid. Since electrical stimulation was found to be most reliable near the rhopalia at the bell margin, the electrodes were embedded at the point farthest radially from the center. As they move with the bell, the force required to bend the electrode wires could restrict animal movement. However, this effect is not fully mechanical since experiments with passive electronics showed no statistically significant decrease in margin movement. Lower frequency stimulation experiments could elucidate the impact of stimulation timing on full relaxation. Further work could also test smaller diameter, but more fragile, wires. 
	
The combination of higher wake power, decreased bell margin movement, and higher total metabolism observed during stimulated swimming shows that externally paced contractions alter the coupling between wake fluid flow and bell kinematics. These results suggest that the increased wake power is consistent with increased energetic throughput resulting from higher pulse rates, rather than a change in per-pulse efficiency. Some work with other species shows that total circulation peaks early in the swim stroke \citep{costelloHydrodynamicsVortexGeneration2019}. This 'overpressure' phenomenon has been described in the vortex ring literature \citep{krueger_significance_2003,gemmellCoolYourJets2021}. Thus, the continued contraction follow-through lost due to changes in swimming kinematics may be important for generating directed momentum transport and higher circulation in the wake. However, previous work has consistently found stimulation increases swimming speeds, suggesting that stimulated animals still produce significant hydrodynamic propulsion \citep{xu_low-power_2020,anuszczykElectromechanicalEnhancementLive2024}.
	
	These experiments suggest a tradeoff between fast swimming and energetic efficiency. For \textit{Aurelia aurita}, these results suggest that there exists an approximately linear relationship between pulse frequency and energy consumption after subtracting basal metabolism. This closely corresponds with the increase in pulse rate $\frac{f_{pulse, stim}}{f_{pulse, nat}}=3.1$ suggesting that energy consumption per pulse is relatively constant across pulse rates.
	The remaining variation is likely due to the changes in swimming kinematics caused by electrical stimulation documented in this work. Cost of transport (COT), defined as $\text{COT}=\frac{\text{Mass}}{\text{Power} \cdot \text{Speed}}$ \citep{gemmell_passive_2013}, necessarily increases with electrical stimulation since COT is assumed to scale as $\text{speed}^2$. Determining the exact scaling relationship between swimming speed and energetic efficiency will require additional experiments. These macroscale measurements show that stimulation at 0.5 Hz increased free-swimming energy consumption by a factor of 5.1$\pm$0.4 with an estimated doubling of average velocity. This corresponds to COT increasing by a factor of 2.5, bringing it closer to larger jetting jellyfish such as \textit{Stomolophus meleagris} \citep{larsonCostsTransportScyphomedusa1987}. These measurements are a factor of 7 larger than COT estimates for natural \textit{Aurelia}. This is due to a combination of electrical stimulation and previous work measuring total metabolism and inferring speed instead of simultaneous measurements resulting in an underestimate of true COT \citep{gemmell_passive_2013}. Since we utilized increased pulse rates, this COT suggests that one contributor to natural jellyfish efficiency is less frequent pulse rates, which often correspond to periods of coasting between pulses. Further investigation by conducting metabolic experiments with lower frequency stimulation could clarify the impact of pulse rate on COT. 
	
	We developed a model that captures the macroscale energy consumption trends observed experimentally. This model underestimated the energetic consumption associated with free-swimming animals, potentially due to the induced unsteady flow field. While the majority of the vorticity is concentrated in vortex rings, other wake energy is distributed spatially, highlighting the complicated nature of jellyfish wake flow dynamics \citep{gemmell_control_2015}. Additional work could refine this model by taking into account more detailed wake reconstructions and vortex measurements. Muscle inefficiency during stimulation could also lead to increased energy consumption. Previous work, using a simple drag model, has explained low energy consumption model predictions by assuming a higher basal metabolism than that measured in this study \citep{acuna_faking_2011}. Additional work is needed to quantify the assumptions regarding energy lost to internal muscle dissipation. \textit{Aurelia aurita} jellyfish bells consist largely of viscoelastic transparent mesoglea \citep{demontMechanicsJetPropulsion1988,megillModulusElasticityFibrillincontaining2005}. The elasticity stores energy during contraction that drives bell relaxation \citep{costello_hydrodynamics_2021}. A significant portion of the stored energy is dissipated in the tissue as heat and never produces motion of the animal body or surrounding fluid. This fraction has been measured to be approximately 0.42 for the hydromedusa \textit{Polyorchis penicillatus} \citep{megillModulusElasticityFibrillincontaining2005,demontMechanicsJetPropulsion1988}. This species is generally smaller than our species of inquiry, but both bells have similar stiffness values and behave as soft, viscoelastic gels \citep{megillModulusElasticityFibrillincontaining2005,gambiniMicroMacrorheologyJellyfish2012}, suggesting the energy dissipated within the mesoglea is likely in a similar range of 0.4-0.5 of stored elastic energy. While the amount of elastically stored energy is unknown, estimates suggest passive energy recapture enables each swimming cycle to propel the animal 30\% farther than without passive energy recapture \citep{gemmell_passive_2013}. Using data from previous work, viscoelastic dissipation is estimated to account for roughly 5-20\% of the mechanical energy expended per swimming pulse in \textit{Polyorchis penicillatus}, excluding additional hydrodynamic losses \citep{demontMechanicsJetPropulsion1988,demontMechanicsJetPropulsion1988a}. A similar contribution of viscoelastic dissipation may therefore be expected in \textit{Aurelia aurita}, although this has not been directly measured.
	
	Experimentally measuring the total metabolism of electrically stimulated jellyfish enables further work developing this platform for ocean exploration. Previous work found portable electrical stimulation combined with an exumbrellar ballasting cap increases animal swimming speeds by up to 4.5 times natural swimming speeds \citep{xu_low-power_2020,anuszczykElectromechanicalEnhancementLive2024}. These stimulated animals are capable of carrying scientific payloads of up to 105\% of animal mass in this ballasting cap, enabling the carrying of a variety of scientific sensors to study the biogeochemistry of the ocean \citep{anuszczykElectromechanicalEnhancementLive2024}. Field tests have demonstrated vertical profiling capability in oceanic conditions near the top of the water column \citep{xuFieldTestingBiohybrid2020,rutledge-in-prep}. Our results show that stimulated animals can swim more than 2.55 km or 15,000 body lengths, which could enable deeper deployment as ocean sensors. Since these continuous swimming experiments were performed in the top 6-m of the water column, deeper deployments would be contingent upon experiments to ensure both the animal and electronics can sustain higher pressures. While the increased COT is higher than natural animals, this COT is more than an order of magnitude smaller than a representative AUV, and thus far more efficient than traditional methods of ocean exploration.
	
	Ocean deployments on the scale of minutes have been demonstrated in previous work \citep{anuszczykIncreasingReliabilityVersatility2025,rutledge-in-prep,xuFieldTestingBiohybrid2020}. The implications of this work necessitate future ocean deployments consider increased COT when choosing electrical stimulation frequency. One option to manage energy consumption during a long-term deployment could include implementing a control loop to vary pulse frequency depending on mission scope and state. Lower pulse frequencies could be selected to conserve energy based on mission progress when decreased swimming speeds are appropriate. 
	
The present measurements are subject to several limitations. Scattering and attenuation within jellyfish tissue limits laser transmission, obscuring detailed features on the far side of the volumetric reconstruction. Wake energy and momentum were evaluated within a finite control volume in the near wake, which may not capture the full extent of the vortex structures as they evolve downstream. As a result, the reported energy ratios should be interpreted as relative measures within a consistent measurement domain between different animal treatments. Future work extending the measurement domain would help further constrain these effects. 
	
Although this study focused on \textit{Aurelia aurita}, these measurements suggest that traditional respirometry methods may underestimate total metabolism of other marine species. The outsized observed impact of hydrodynamic drag on total metabolism may be underrepresented in studies relying on confined experimental configurations. The techniques utilized here hold potential for application to other marine species. Application of 3D-3C scanning PIV to wake analysis may reveal additional wake dynamics and could be used to explore the relationships between biomechanics and fluid flow. Recent work has achieved detailed laser scans of \textit{Nanomia bijuga} and \textit{Cystisoma} \citep{danielsNewMethodRapid2023}. Our work extends these findings with automated reconstructions of larger transparent animals. Reconstructions could be utilized for comparative studies of animal morphology, biomechanics, and behavior given sufficient camera frame rates to limit movement between successive laser scans. This catabolysis metabolism technique could be extended to a broad range of gelatinous zooplankton but would require modification for application to more complex organisms. While jellyfish have little tissue type variation, application to some vertebrates, e.g. bony fishes, would require tracking tissue type lost through catabolysis to account for differing chemistry and carbon content.
	
	\section{Acknowledgments}
	The authors would like to thank Cabrillo Marine Aquarium and Aquarium of the Pacific for providing jellyfish used in these experiments. The authors also acknowledge helpful conversations with John Costello, Sean Colin, Brad Gemmell, Lea Goentoro, Michael Dickinson, Morteza Gharib, Emily Carrington, and Mark Denny. The authors would like to thank Marni Rosenthal for assistance in designing Figure \ref{vert}A. Matt Fu provided valuable assistance with the laser scanning setup.
	
	\section{Competing interest}
	The authors declare that they have no competing interests.
	
	\section{Funding}
	This work was supported by the National Science Foundation Graduate Research Fellowship Grant Number DGE-1745301, National Science Foundation Grant 2311867, funding from the U.S. Office of Naval Research, and from Friday Harbor Laboratories.
	
	\section{Author contributions}
	S.R.A. and J.O.D. conceived of project, S.R.A. and K.P. conducted experiments, S.R.A. and J.O.D. analyzed results, S.R.A. wrote initial draft and all authors edited paper.
	
	\section{Data availability}
	Raw image data available upon request.
	
	\section{Ethics statement}
	All experiments were conducted in accordance with institutional guidelines for the care and use of animals. Because \textit{Aurelia aurita} are invertebrates, formal IACUC approval was not required, but all procedures followed established ethical standards for minimizing harm.
	
\begingroup
\setlength{\itemsep}{0pt}
\setlength{\parskip}{0pt}
\setlength{\parsep}{0pt}
\setlength{\baselineskip}{10.5pt}

\endgroup
	
\begin{appendices}

\newpage
\section{Supplementary info}

\subsection{Supplementary methods}

\subsubsection{Laser scanning PIV}
PIV data of the swimming jellyfish was taken using a three-dimensional, three-component single-camera laser scanning system, shown in Figure \ref{scanning}, first applied to swimming shrimp \citep{fu_single-camera_2021}. This system used a 671 nm continuous laser (5 W Laserglow LRS0671 DPSS Laser System) and a series of optics to generate a scanning laser sheet. The optics consisted of a mirror mounted on a rotating galvanometer (Thorlabs GVS211/M, 10 kHz bandwidth) controlled with an arbitrary function generator (Tektronix AFG3011C) to sweep across the volume of inquiry at 10 Hz; a 25 cm diameter condenser lens to redirect the laser normal to the scanning direction; and a cylindrical, 16-mm diameter glass rod used to form the laser beam into a thin sheet. The laser sheet was set to sweep normal to its plane across a distance of 18.5 cm. A high-speed camera recording at 8000 fps (Photron FASTCAM SA-Z) thus captured 800 images at a resolution of 1024 x 1024 pixels, or 22.3 x 22.3 cm, for each volumetric scan. The jellyfish were manually directed downwards through the scanning volume in a 1.2 m tall x 0.3 m x 0.3 m tank filled with seawater at 35 ppt. The fluid was seeded with 20 \textmu m neutrally-buoyant polyamide particles for flow visualization (LaVision Inc., Ypsilanti, MI). This experimental scanning setup was calibrated using a 3D-printed calibration plate with a grid of through holes spaced every 1 cm. This plate was mounted at 45 degrees to the camera. The calibration plate was scanned, reconstructed, and the hole-center to hole-center distance was measured for the entire grid. A rotational transformation was applied to these spacings, recovering the three-dimensional calibration.
	
Because the flow field evolves much more slowly than the temporal resolution of the measurements, the measurements across a volume can be treated as quasi-instantaneous. Volumes are recorded every 0.1 s, while the jellyfish typically swims at 1-3 cm s$^{-1}$. Thus, over a single sampling interval, the animal translates only 0.1-0.3 cm, which is small compared with the characteristic size of the bell and the wake structures being measured. As a result, the velocity field and associated metrics change only slightly across a measurement, allowing each volume to be treated as an instantaneous snapshot of a slowly evolving flow \citep{adrianParticleImageVelocimetry2011}. Under these conditions, treating the measurements as such provides a reasonable approximation for this analysis.
        
		Table 1 includes the volumetric PIV parameters used for microscale wake measurements.
	
		\begin{table}[!h]
			\centering
			\caption{\textbf{Volumetric PIV acquisition and processing parameters.}}
			\label{tab:piv_params}
			
			\begin{tabularx}{\columnwidth}{l X}
				\toprule
				
				\multicolumn{2}{l}{\textbf{Acquisition}} \\
				\midrule
				Raw image resolution & $1024 \times 1024$ pixels \\
				Scan rate & \SI{10}{Hz} \\
				Time between volumes ($\Delta t$) & \SI{0.1}{s} \\
				Frames per volume & 800 \\
				Volumes per trial & 105 \\
				Vector fields per trial & 104 \\
				Voxel spacing ($x,y$) & \SI{0.0217}{cm} \\
				Voxel spacing ($z$) & \SI{0.0232}{cm} \\
				
				\midrule
				\multicolumn{2}{l}{\textbf{PIV processing}} \\
				\midrule
				Interrogation window size & $32 \times 32 \times 32$ voxels \\
				Window overlap & $16 \times 16 \times 16$ voxels \\
				Search area size & $32 \times 32 \times 32$ voxels \\
				Vector spacing & 16 voxels ($\approx$ \SI{0.35}{cm} $\times$ \SI{0.35}{cm} $\times$ \SI{0.37}{cm}) \\
				Final vector grid size & $\sim 63 \times 52 \times 49$ vectors (crop-dependent) \\
				Subpixel peak estimation & Gaussian fit \\
				Signal-to-noise metric & Peak-to-peak \\
				Mask dilation & $\approx$ half interrogation window size \\
				
				\midrule
				\multicolumn{2}{l}{\textbf{Vector validation and filtering}} \\
				\midrule
				Vector validation threshold & SNR $\ge 1$ \\
				Velocity outlier cutoff & $|u| >$ \SI{15}{cm\,s^{-1}} removed \\
				Velocity smoothing & 3D Gaussian filter ($\sigma = 2$ voxels) \\
				Additional smoothing & $5 \times 5 \times 5$ mean filter \\
				
			\end{tabularx}
			
		\end{table}
        
\subsubsection{Post-processing}
A custom Python code was written to mask the three-dimensional jellyfish out of each image stack. The code identified the jellyfish and performed binarization, thresholding, and morphological opening and closing operations to create a detailed reconstruction of the animal, as shown in Figure \ref{scanning}C. After masking, the open source Python-based tool OpenPIV was used to process the image stacks and generate vectors \citep{liberzonOpenPIVOpenpivpythonFixed2019}. Velocities were computed by correlating consecutive volumetric scans, yielding one 3D velocity field per 0.1 s. Because the volume is acquired sequentially, each 3D field represents a quasi-instantaneous volume. MATLAB was used for filtering, data visualization, and analysis, and all results use the post-processed field. Velocity magnitudes exceeding \SI{15}{cm/s^{-1}} were removed as spurious outliers, as these values are more than an order of magnitude larger than the observed swimming speeds and expected flow velocities for jellyfish pulses of these diameters.

\subsubsection{Basal metabolism}
The total metabolism of free-swimming animals comprises both the metabolic requirements to maintain swimming and basal metabolism for other physiological processes. Additional experiments were conducted to quantify the basal metabolic component of the total metabolic measurements. To the best of our knowledge, these are the first measurements of the basal metabolism of jellyfish without animal movement. Basal metabolic experiments were performed using a 2.55 liter glass respirometry chamber and the oxygen probe and temperature sensor described in this previous section with 6 animals. Two methods were investigated successively: anesthetization with magnesium chloride (MgCl$_2$) to suppress muscle movement and excision of the nerve centers, or rhopalia, of the animal in order to arrest endogenous swimming signals. Excising rhopalia has been shown to arrest normal contractions \citep{satterlie_neuronal_2002}, and the animals were observed to not contract for the duration of this experiment after excision. Each animal was transferred to the respirometry chamber filled with artificial seawater, where they underwent five distinct treatments of 1 hour each. This enabled testing of both methods and comparison with natural respiration rates. First, the chamber was closed, and the normal respiration rate was measured. Next, the animal was removed and MgCl$_2$ (AC413410025, Fisher Scientific, Waltham, MA, USA) was added at a molarity of 0.3 M. The MgCl$_2$ was mixed for 5 minutes until dissolution, and the animal was added for the anesthetization test. Upon completion, the water was then exchanged for clean artificial seawater, and a second normal respiration rate was measured to ensure the animal had fully recovered from anesthetization. The same animals were then used to test the excision method. Respiratory rate was measured again for the immobilized jellyfish after excision. The animal was again removed from the tank, the water was replaced with the MgCl$_2$ solution, and respiratory rate was measured combining both anesthetization and excision methods. While both anesthetization and excision methods were tested, we report results from the excision method here due to potential confounding impacts of MgCl$_2$ with animal tissue on oxygen consumption. 

\subsubsection{Laser scanning for volumetric reconstruction}
Volumetric reconstructions of the jellyfish were created using the same single-camera laser scanning system as described in Figure \ref{scanning}. Optical parameters were changed slightly. The laser sheet swept across a distance of 11.7 cm, and camera recorded at 6000 fps thus capturing 600 1024 x 850 pixel or 22.5 x 18.7 cm 2D images for each scan. A similar code was used to generate 3D masks of the jellyfish. 

\subsubsection{Energetic consumption from volumetric changes}
In order to validate this conversion, simultaneous laser scanning and closed system respirometry experiments were performed. The jellyfish was scanned and reconstructed using the procedure described in this section and immediately transferred to a 15.1  liter closed respirometry chamber filled with artificial seawater at 35 ppt (Fig. 1C). An oxygen probe (OPTO-430, Unisense, Denmark) and temperature sensor (TEMP-UNIAMP-7, Unisense, Denmark) were used to continuously measure the oxygen concentration at a sampling frequency of 0.1 Hz. After 24 hours, the animal was removed, scanned, and the artificial seawater in the respirometry chamber was flushed and replaced. The animal was returned to the respirometry chamber for a second 24 hour respirometry measurement. Finally, at 48 hours from experiment initiation, the animal was removed, scanned, and the experiment was completed. The respirometry chamber was chosen to be large enough that the animal avoided hypoxic conditions over the course of the experiment. This experiment was repeated for 3 animals. Oxygen consumption in an empty respirometry chamber was measured, and background oxygen consumption was subtracted from experimental results. A linear fit to the data was used as the average respiratory rate (RR). These values were compared with literature values for \textit{Aurelia aurita} wet weight (WW) given by 
	
	\begin{align}\label{1.1}
		RR=0.0765WW^{1.038}
	\end{align}
	
	\citep{uyePopulationBiomassFeeding2005}.

\subsubsection{Carbon content}
Tissue carbon content, $C$, was experimentally measured for 9 animals to measure body composition changes while undergoing catabolysis. The animals were transferred to a clean 108 liter artificial seawater tank without food. Three animals were randomly selected for removal at each timestep of 0 hours, 25 hours, and 50 hours after the experiment start. Each animal was patted dry, $WW$ measured on an electric balance, and transferred to an electric oven where they were dried at 60-65°C until a constant weight was achieved. Dried samples were pulverized in a mortar and pestle and weighed. Samples were then analyzed for carbon and nitrogen content on an elemental analyzer (Santa Barbara Marine Science Institute Analytical Lab, CEC 440HA, Control Equipment Corp., MA, USA). These experiments did not show statistically significant changes in carbon content over 50 hours without feeding. We found an average carbon content of 0.105\% ± 0.026\% of $WW$, which is not statistically different from the 0.13\% ± 0.06\% reported in the literature \citep{uyePopulationBiomassFeeding2005}. We used the value measured in the present experiments for the laser scanning method calculations. Analysis showed an average carbon-to-nitrogen ratio of 3.64 $\pm$ 1.26 which is not statistically different from the literature value of 3.71 $\pm$ 1.63 \citep{uyePopulationBiomassFeeding2005}.

\subsubsection{Macroscale stimulated free-swimming experiments}
	
The laser scanning method was utilized to measure the free-swimming energy consumption of electrically stimulated, continuous swimming jellyfish in a 6-m tall, 13,600-liter vertical tank (Figure \ref{vert}A). The tank was filled with artificial seawater balanced at 35 ppt and 21°C and has a continuous filtration pump, which was turned off at least 2 hours before these experiments. This tank is divided into a test section on the right, where experiments are conducted, and a recirculation region on the left with water pumps (A1200, Abyzz, Germany). This enabled continuous flow, shown with red arrows, through the test section functioning as a “flow treadmill.” During long-duration swimming experiments, a jellyfish was equipped with stimulation electronics, weighted to maintain slight positive buoyancy with a stable downward orientation, and released at the top of the tank to swim downwards. A camera microprocessor (RT1062 Cam, OpenMV, Atlanta, GA, USA), set back 0.86 m from the tank, was programmed with a custom MicroPython computer vision algorithm. This computer vision camera tracked the animal location during continuous swimming and was programmed to interface with the water pumps while running a proportional-integral-derivative control loop. Figure \ref{vert}B shows the view from the camera of the middle section of the tank as a jellyfish was swimming downwards. As the animal descends in the tank, the controller continuously adjusts the flow opposing the jellyfish with an optimization goal of keeping the animal within the middle section of the tank and in view of the camera. The computer vision camera was programmed to track both the animal location and the pump power. Particle Image Velocimetry (PIV) was conducted to calibrate pump power to pump flow speed and confirm flow uniformity. Animal swimming speed was found by combining the movement of the animal in the lab frame with the pump flow speed. During the course of the 50 hour experiments, the animal was occasionally out of frame for brief periods. These periods were removed when calculating swimming metrics for speed and distance. 

\subsection{Supplementary results}
		\begin{figure} [!htbp]
			\centering\includegraphics[width=0.5\textwidth]{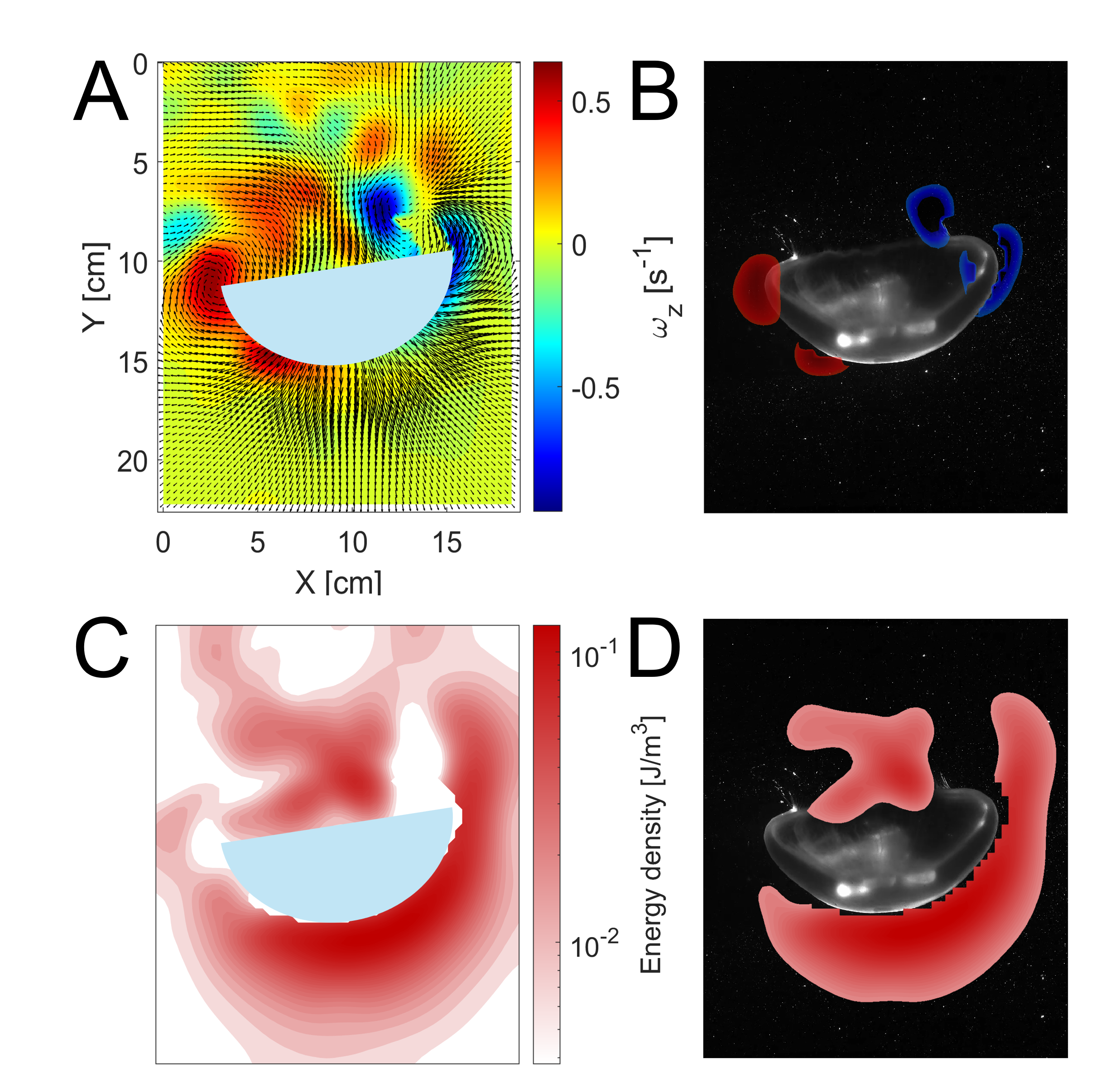}
			\caption{\textbf{Contracted jellyfish PIV flow visualization.} (A) Out-of-plane 2D vorticity visualization with overlaid velocity vector field. Jellyfish cartoon mask is shown for illustration. Color bar shows regions of positive and negative vorticity and vorticity cores are visible at margin of jellyfish bell and in the wake. (B) Vorticity with same color scale overlaid on raw image of jellyfish at center plane illuminated by laser. Regions of low magnitude vorticity are filtered out. (C) Kinetic energy density map illustrates regions of high energy in wake with fully contracted jellyfish. Energy density in units of joules/m$^3$. Cartoon jellyfish mask is shown for illustration and color bar is in log scale. (D) Kinetic energy density map with the same color scale overlaid on raw image of jellyfish at center plane illuminated by laser. Regions of low energy density are filtered out for visualization.}
			\label{flowviz_contract}
		\end{figure}

		\begin{figure}[!htbp]
			\centering\includegraphics[width=0.5\textwidth]{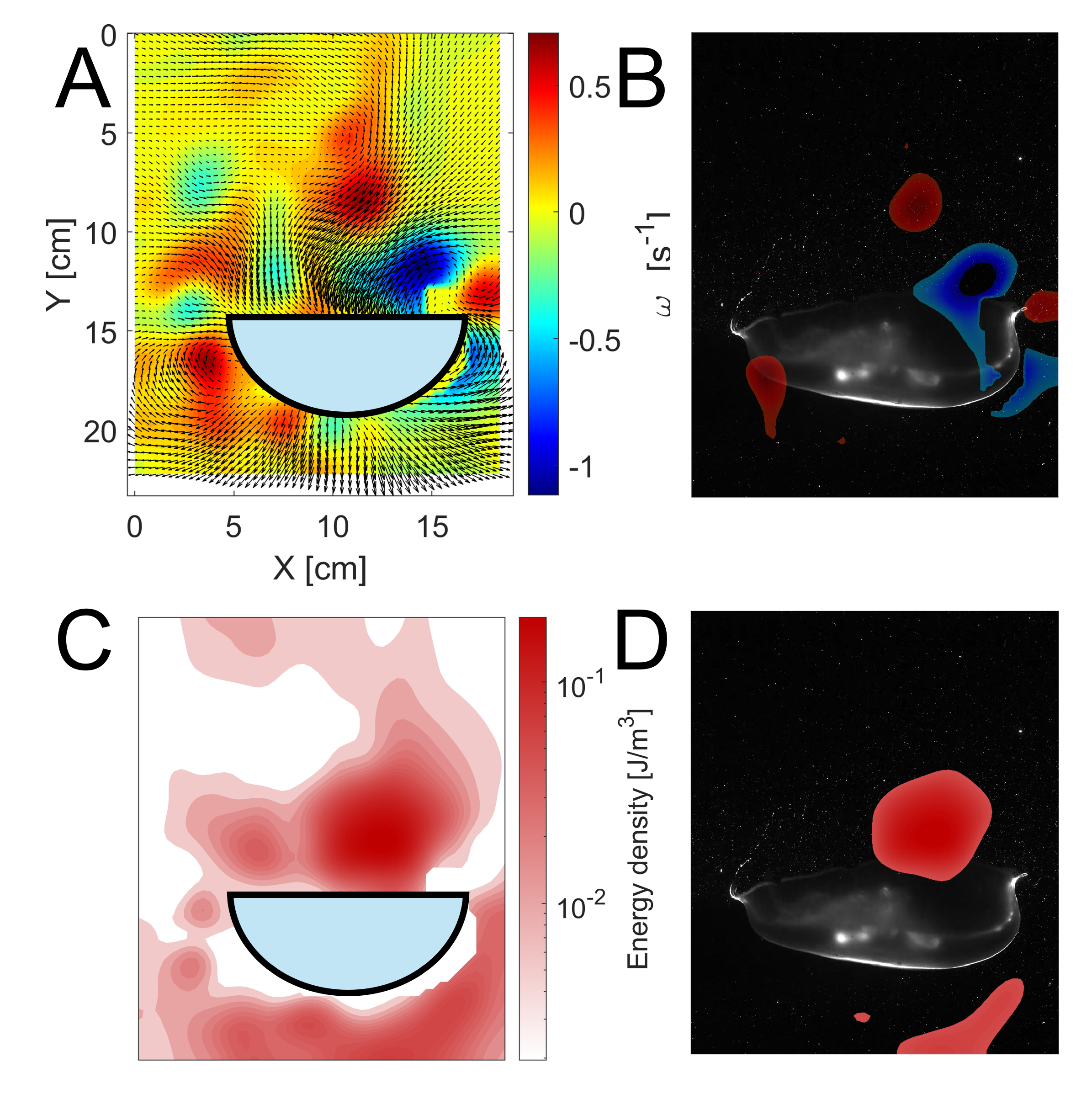}
			\caption{\textbf{Relaxed jellyfish PIV flow visualization}  (A) Out-of-plane 2D vorticity visualization with overlaid velocity vector field. Jellyfish cartoon mask is shown for illustration. Color bar shows regions of positive and negative vorticity and vorticity cores are visible at margin of jellyfish bell and in the wake. (B) Vorticity with same color scale overlaid on raw image of jellyfish at center plane illuminated by laser. Regions of low magnitude vorticity are filtered out. (C) Kinetic energy density map illustrates regions of high energy in wake with fully relaxed jellyfish. Energy density in units of joules/m$^3$. Cartoon jellyfish mask is shown for illustration and color bar is in log scale. (D) Kinetic energy density map with the same color scale overlaid on raw image of jellyfish at center plane illuminated by laser. Regions of low energy density are filtered out for visualization.}
			\label{flowviz_relax}
		\end{figure}
		
		To validate the microscale wake dynamics measurements, we visualized the vortex rings at the center plane. Figures \ref{flowviz_contract} and \ref{flowviz_relax} show 2D visualizations of the flow fields for a single representative jellyfish at different points in time. Figure \ref{flowviz_contract} illustrates the flow fields around a fully contracted jellyfish bell, while Figure \ref{flowviz_relax} shows a fully relaxed bell. Figure \ref{flowviz_contract}A shows the 2D vorticity $\omega_z$ for a representative experiment.
		
		A cartoon jellyfish mask is shown, and opposite rotating vortex ring cores are clearly visible. Figure \ref{flowviz_contract}B shows the same vorticity field filtered and overlaid on top of the raw still image at the same location and time. As the jellyfish pulses, it releases vortex rings, and additional vorticity builds on the outside of the bell. To illustrate kinetic energy, Figure \ref{flowviz_contract}C shows the 2D kinetic energy density map in units of joules/m$^3$ at the same location in time and space with the cartoon jellyfish mask. Figure \ref{flowviz_contract}D shows the same energy density map filtered and overlaid on top of the raw jellyfish image. This colorbar is shown with a log scale. 
		
		Comparing the energy density in  Figures \ref{flowviz_contract}C-D and \ref{flowviz_relax}C-D illustrates how these quantities vary over time. During contraction, energy density peaks in the bow wave but then falls as the animal slows during relaxation, when energy density peaks in the trailing wake behind the animal.

		\begin{sidewaysfigure}[p]
			\centering
			\includegraphics[width=0.5\textwidth]{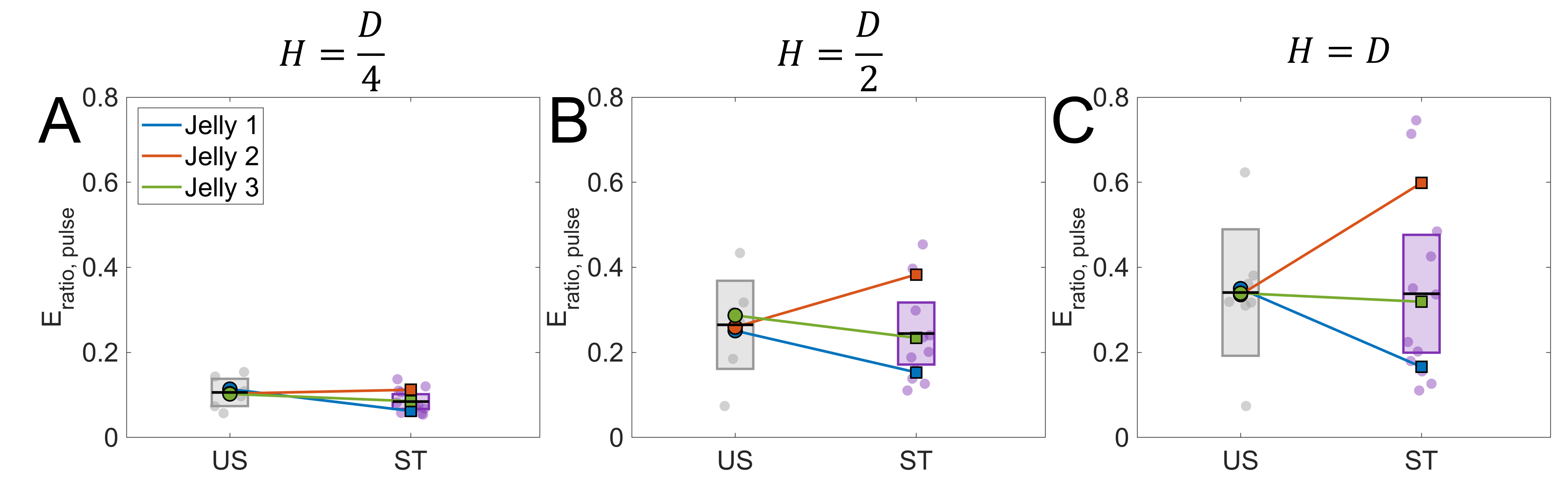}
			\caption{\textbf{Sensitivity of normalized wake energy to control-volume height.} Sensitivity of the pulse-averaged wake energy ratio to the height of the cylindrical control volume. Panels show results for control volume heights of (A) $H=\frac{D}{4}$, (B) $H=\frac{D}{2}$, and (C) $H=D$, where $D$ is the bell diameter. Grey and purple points denote individual pulses for unstimulated (US) and stimulated (ST) trials, respectively. Colored symbols connected by lines indicate animal means. Shaded boxes show the 95\% confidence interval of the pooled pulse mean for each condition. The primary analysis uses the $H=\frac{D}{2}$ control volume shown in panel (B).}
			\label{sense}
		\end{sidewaysfigure}
		
		The primary analysis used a control volume height of $H=\frac{D}{2}$. Repeating the analysis with smaller and larger control volumes ($H=\frac{D}{4}$ and $H=D$) produces similar qualitative patterns across animals and conditions (Figure \ref{sense}). This indicates that the conclusions are not sensitive to moderate changes in control-volume height.

\subsection{Macroscale stimulated free-swimming experiments}
	
	\begin{figure*}[!h]
		\centering\includegraphics[width=1\textwidth]{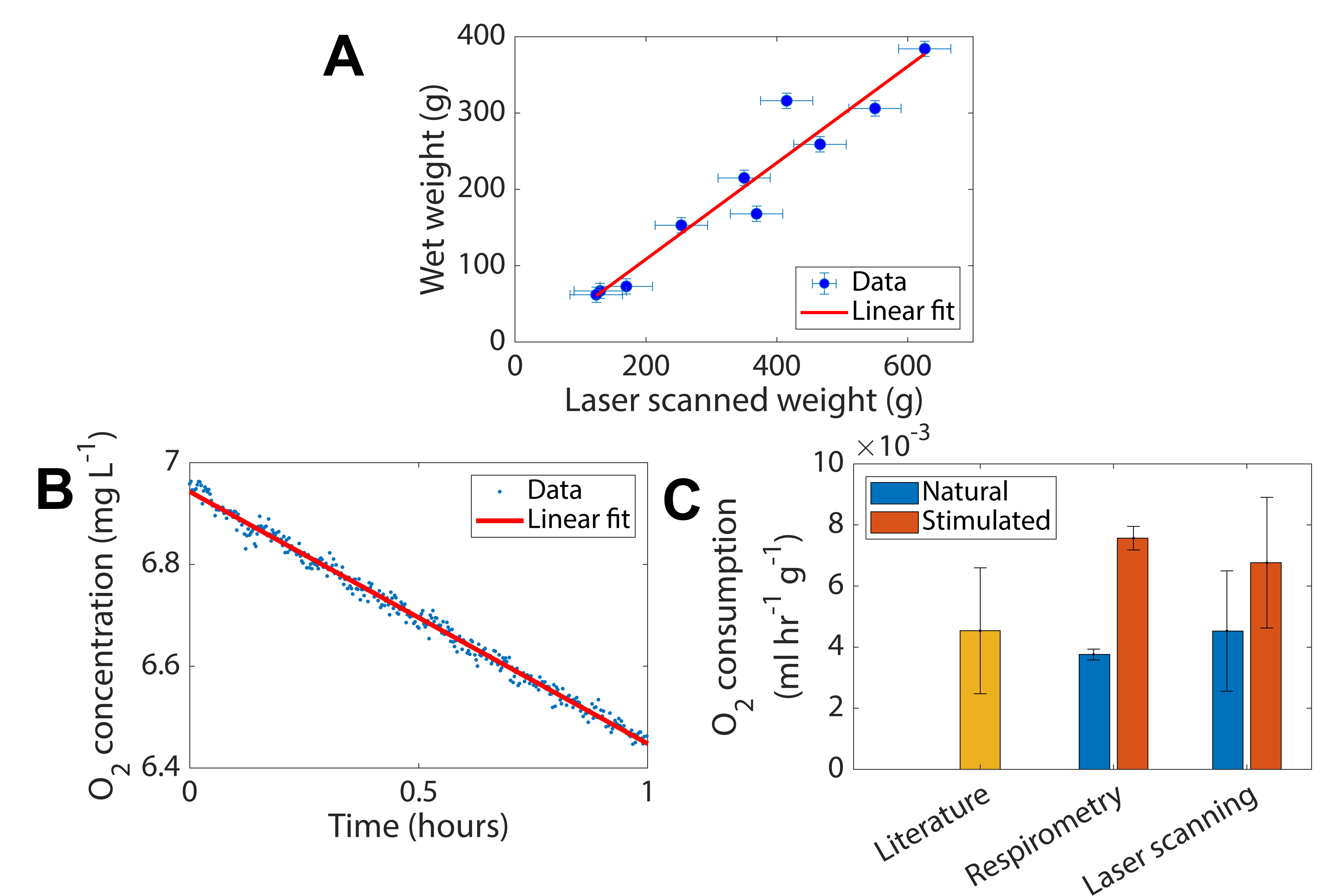}
		\caption{\textbf{Validation of laser scanning technique for energy measurements.} (A) Linear relationship between wet weight (WW) and laser scanned weight (LSW) of $WW=0.63LSW-17.66$ for 10 animals. (B) A sample 1 hour closed system respirometry experiment with linear fit and R-squared value of $R^2=0.9921$. This high R-squared value confirms the linearity of jellyfish oxygen consumption over 1 hour. (C) Literature values for natural animals compared with simultaneous closed system respirometry and laser scanning. There is no statistical difference between the oxygen consumption rates found with the two different techniques. Oxygen consumption is given in terms of ml O$_2$ per hour per g wet weight. All error bars represent 1 standard deviation.}
		\label{combo}
	\end{figure*}
	
\subsubsection{Laser scanning technique validation}
We compared wet weight ($WW$) to laser scanned weight ($LSW$) to validate this technique. Figure \ref{combo}A shows the linear relationship between the measured $WW$ and $LSW$ of $WW=0.63LSW-17.66$ with $R^2=0.9256$ for 10 animals of different sizes. The $LSW$ overestimated the $WW$ of the animal, likely due to the scattering of the laser sheet through the tissue. This well-defined relationship enabled us to convert accurately between $WW$ and $LSW$. 
	
Simultaneous laser scanning and respirometry experiments confirmed the accuracy of this method. Figure \ref{combo}B shows a representative 1 hour respirometry experiment. A linear fit to this data has an R-squared value of $R^2=0.9921$ and illustrates the linear trend in oxygen consumption over the course of these experiments. The literature bar represents the wet weight normalized oxygen consumption for an animal of the same size as those used in the respirometry and laser scanning experiments shown in Figure \ref{combo}C. This was found using Equation \ref{1.1} with the error given as ± 1 standard deviation. The respirometry bar compares the animal’s natural and electrically stimulated respiratory rate we measured in the respirometry chamber with the error given by ± 1 standard deviation of the linear fit to the respirometry data. The laser scanning bar in Figure \ref{combo}C also compares naturally swimming animals to electrically stimulated animals measured with the laser scanning method. This error represents ± 1 standard deviation and combines the errors in scanning and Equation 3.3. Electrically stimulated animals were found on average to consume 73\% ± 32\% more energy than naturally swimming animals in this confined configuration.
	
These experiments validate that the observed laser scanning respiratory rate from catabolysis, calculated using Equation \ref{RR_ikeda}, corresponds to measurements from the respirometry chamber. The blue “Natural” bars have no statistically significant difference with a two-tailed $t$-test p-value of 0.44. Similarly, the red “Stimulated” bars have no statistically significant difference with a two-tailed $t$-test p-value of p=0.60. This confirms that although the laser scanning method has a higher error compared to the respirometry chamber, the two methods are in agreement and the technique is suitable for further experiments.

\end{appendices}


\begin{thebibliography}{10}

\bibitem{degrootIncorporatingOtolithisotopeInferred2024}
Valesca~A de~Groot, Clive Trueman, and Amanda~E Bates.
\newblock Incorporating otolith-isotope inferred field metabolic rate into conservation strategies.
\newblock 12(1):coae013.

\bibitem{vogelLifeMovingFluids1994}
Steven Vogel.
\newblock {\em Life in Moving Fluids: The Physical Biology of Flow - Revised and Expanded Second Edition}.
\newblock Princeton University Press, {NED} - new edition edition.

\bibitem{dabiri_wake-based_2010}
J.~O. Dabiri, S.~P. Colin, K.~Katija, and J.~H. Costello.
\newblock A wake-based correlate of swimming performance and foraging behavior in seven co-occurring jellyfish species.
\newblock 213(8):1217--1225.

\bibitem{lighthillMathematicalBiofluiddynamics1997}
M.~J. Lighthill.
\newblock {\em Mathematical biofluiddynamics}.
\newblock Society for Industrial and Applied Mathematics.

\bibitem{costello_hydrodynamics_2021}
John~H. Costello, Sean~P. Colin, John~O. Dabiri, Brad~J. Gemmell, Kelsey~N. Lucas, and Kelly~R. Sutherland.
\newblock The hydrodynamics of jellyfish swimming.
\newblock 13(1):375--396.
\newblock \_eprint: https://doi.org/10.1146/annurev-marine-031120-091442.

\bibitem{elsingaTomographicParticleImage2006}
G.~E. Elsinga, F.~Scarano, B.~Wieneke, and B.~W. van Oudheusden.
\newblock Tomographic particle image velocimetry.
\newblock 41(6):933--947.

\bibitem{adrianParticleImageVelocimetry2011}
Ronald~J. Adrian and Jerry Westerweel.
\newblock {\em Particle Image Velocimetry}.
\newblock Cambridge University Press.

\bibitem{fu_single-camera_2021}
Matt~K. Fu, Isabel~A. Houghton, and John~O. Dabiri.
\newblock A single-camera, 3d scanning velocimetry system for quantifying active particle aggregations.
\newblock 62(8):168.

\bibitem{gemmell_control_2015}
Brad~J. Gemmell, Daniel~R. Troolin, John~H. Costello, Sean~P. Colin, and Richard~A. Satterlie.
\newblock Control of vortex rings for manoeuvrability.
\newblock 12(108):20150389.

\bibitem{mohebbiMeasurementsModellingInduced2024}
Nina Mohebbi, Joonha Hwang, Matthew~K. Fu, and John~O. Dabiri.
\newblock Measurements and modelling of induced flow in collective vertical migration.
\newblock 1001:A50.

\bibitem{svendsenDesignSetupIntermittentflow2016}
M.~B.~S. Svendsen, P.~G. Bushnell, and J.~F. Steffensen.
\newblock Design and setup of intermittent-flow respirometry system for aquatic organisms.
\newblock 88(1):26--50.
\newblock \_eprint: https://onlinelibrary.wiley.com/doi/pdf/10.1111/jfb.12797.

\bibitem{harrisICESZooplanktonMethodology2000}
T.~Ikeda, J.J. Torres, S.~Hernandez-Leon, S.P. Geiger, Roger Harris, Peter Wiebe, Jurgen Lenz, Hein-Rune Skjoldal, and Mark Huntley.
\newblock {\em {ICES} Zooplankton Methodology Manual}.
\newblock Elsevier.

\bibitem{trebergEstimatesMetabolicRate2016}
Jason~R. Treberg, Shaun~S. Killen, Tyson~J. {MacCormack}, Simon~G. Lamarre, and Eva~C. Enders.
\newblock Estimates of metabolic rate and major constituents of metabolic demand in fishes under field conditions: Methods, proxies, and new perspectives.
\newblock 202:10--22.

\bibitem{simpkinsPhysiologicalResponsesJuvenile2003}
Darin~G. Simpkins, Wayne~A. Hubert, Carlos~Martinez Del~Rio, and Daniel~C. Rule.
\newblock Physiological responses of juvenile rainbow trout to fasting and swimming activity: effects on body composition and condition indices.
\newblock 132(3):576--589.

\bibitem{lilleyIndividualShrinkingEnhance2014}
Martin~{KS} Lilley, Amanda Elineau, Martina Ferraris, Alain Thiéry, Lars Stemmann, Gabriel Gorsky, and Fabien Lombard.
\newblock Individual shrinking to enhance population survival: quantifying the reproductive and metabolic expenditures of a starving jellyfish, pelagia noctiluca.
\newblock 36(6):1585--1597.

\bibitem{ishiiFoodRegulationGrowth1998}
Haruto Ishii and Ulf Båmstedt.
\newblock Food regulation of growth and maturation in a natural population of aurelia aurita (l.).
\newblock 20(5):805--816.

\bibitem{fuBodySizeReduction2014}
Zhilu Fu, Masashi Shibata, Ryosuke Makabe, Hideki Ikeda, and Shin-ichi Uye.
\newblock Body size reduction under starvation, and the point of no return, in ephyrae of the moon jellyfish aurelia aurita.
\newblock 510:255--263.

\bibitem{bondyale-juez_wind_2022}
Daniel~R. Bondyale-Juez, Vanesa Romero-Kutzner, Jennifer~E. Purcell, Ico Martínez, Theodore~T. Packard, and May Gómez.
\newblock Wind drifting vs. pulsating swimming jellyfish: Respiratory metabolism and composition differences in physalis physalis, velella velella, aurelia aurita, and pelagia noctiluca.
\newblock 9.

\bibitem{danielsNewMethodRapid2023}
Joost Daniels, Giovanna Sainz, and Kakani Katija.
\newblock New method for rapid 3d reconstruction of semi-transparent underwater animals and structures.
\newblock 5(1):obad023.

\bibitem{harrisICESZooplanktonMethodology1999}
Roger Harris.
\newblock {\em {ICES} zooplankton methodology manual}.
\newblock Academic.

\bibitem{tillsReducedPHAffects2016}
O.~Tills, X.~Sun, S.~D. Rundle, T.~Heimbach, T.~Gibson, A.~Cartwright, M.~Palmer, T.~Rudin-Bitterli, and J.~I. Spicer.
\newblock Reduced {pH} affects pulsing behaviour and body size in ephyrae of the moon jellyfish, \textit{Aurelia aurita}.
\newblock 480:54--61.

\bibitem{dillonEffectsAcuteChanges1977}
T.~M. Dillon.
\newblock Effects of acute changes in temperature and salinity on pulsation rates in ephyrae of the scyphozoan aurelia aurita.
\newblock 42(1):31--35.

\bibitem{mchenryOntogeneticScalingHydrodynamics2003}
Matthew~J. {McHenry} and Jason Jed.
\newblock The ontogenetic scaling of hydrodynamics and swimming performance in jellyfish (aurelia aurita).
\newblock 206(22):4125--4137.

\bibitem{xu_low-power_2020}
Nicole~W. Xu and John~O. Dabiri.
\newblock Low-power microelectronics embedded in live jellyfish enhance propulsion.
\newblock 6(5):eaaz3194.

\bibitem{anuszczykElectromechanicalEnhancementLive2024}
Simon~R. Anuszczyk and John~O. Dabiri.
\newblock Electromechanical enhancement of live jellyfish for ocean exploration.
\newblock 19(2):026018.

\bibitem{ruizVortexenhancedPropulsion2011}
Lydia~A. Ruiz, Robert~W. Whittlesey, and John~O. Dabiri.
\newblock Vortex-enhanced propulsion.
\newblock 668:5--32.

\bibitem{li_volumetric_2023}
Derek~J. Li and Leah Mendelson.
\newblock Volumetric measurements of wake impulse and kinetic energy for evaluating swimming performance.
\newblock 64(3):47.

\bibitem{katijaViscosityenhancedMechanismBiogenic2009}
Kakani Katija and John~O. Dabiri.
\newblock A viscosity-enhanced mechanism for biogenic ocean mixing.
\newblock 460(7255):624--626.

\bibitem{hedrick_software_2008}
Tyson~L Hedrick.
\newblock Software techniques for two- and three-dimensional kinematic measurements of biological and biomimetic systems.
\newblock 3(3):034001.

\bibitem{saffmanVortexDynamics1993}
P.~G. Saffman.
\newblock {\em Vortex Dynamics}.
\newblock Cambridge Monographs on Mechanics. Cambridge University Press.

\bibitem{rosenheadLaminarBoundaryLayers1963}
Louis Rosenhead.
\newblock {\em Laminar boundary layers; an account of the development, structure, and stability of laminar boundary layers in incompressible fluids, together with a description of the associated experimental techniques}.
\newblock Oxford [Eng.] Clarendon Press.

\bibitem{batchelorIntroductionFluidDynamics2000}
G.~K. Batchelor.
\newblock {\em An Introduction to Fluid Dynamics}.
\newblock Cambridge Mathematical Library. Cambridge University Press.

\bibitem{uyePopulationBiomassFeeding2005}
S.~Uye and H.~Shimauchi.
\newblock Population biomass, feeding, respiration and growth rates, and carbon budget of the scyphomedusa aurelia aurita in the inland sea of japan.
\newblock 27(3):237--248.

\bibitem{frandsenSizeDependentRespiration1997}
Kristian~Toft Frandsen and Hans~Ulrik Riisgård.
\newblock Size dependent respiration and growth of jellyfish, aurelia aurita.
\newblock 82(4):307--312.
\newblock \_eprint: https://doi.org/10.1080/00364827.1997.10413659.

\bibitem{dabiriFastswimmingHydromedusaeExploit2006}
John~O. Dabiri, Sean~P. Colin, and John~H. Costello.
\newblock Fast-swimming hydromedusae exploit velar kinematics to form an optimal vortex wake.
\newblock 209(11):2025--2033.

\bibitem{acuna_faking_2011}
José~Luis Acuña, Ángel López-Urrutia, and Sean Colin.
\newblock Faking giants: The evolution of high prey clearance rates in jellyfishes.
\newblock 333(6049):1627--1629.

\bibitem{gemmell_passive_2013}
Brad~J. Gemmell, John~H. Costello, Sean~P. Colin, Colin~J. Stewart, John~O. Dabiri, Danesh Tafti, and Shashank Priya.
\newblock Passive energy recapture in jellyfish contributes to propulsive advantage over other metazoans.
\newblock 110(44):17904--17909.

\bibitem{lighthillFundamentalsConcerningWave1986}
James Lighthill.
\newblock Fundamentals concerning wave loading on offshore structures.
\newblock 173:667--681.

\bibitem{hoernerFluidDynamicDrag}
Sighard~F Hoerner.
\newblock Fluid-dynamic drag.

\bibitem{garciaOxygenSolubilitySeawater1992}
Herncin~E. Garcia and Louis~I. Gordon.
\newblock Oxygen solubility in seawater: Better fitting equations.
\newblock 37(6):1307--1312.

\bibitem{Daniel1983MechanicsEnergeticsMedusanb}
Thomas Daniel.
\newblock Mechanics and energetics of medusan jet propulsion.

\bibitem{larson_respiration_1987}
R.~J Larson.
\newblock Respiration and carbon turnover rates of medusae from the {NE} pacific.
\newblock 87(1):93--100.

\bibitem{costelloHydrodynamicsVortexGeneration2019}
John~H. Costello, Sean~P. Colin, Brad~J. Gemmell, and John~O. Dabiri.
\newblock Hydrodynamics of vortex generation during bell contraction by the hydromedusa eutonina indicans (romanes, 1876).
\newblock 4(3):44.

\bibitem{krueger_significance_2003}
Paul~S. Krueger and M.~Gharib.
\newblock The significance of vortex ring formation to the impulse and thrust of a starting jet.
\newblock 15(5):1271--1281.

\bibitem{gemmellCoolYourJets2021}
Brad~J. Gemmell, John~O. Dabiri, Sean~P. Colin, John~H. Costello, James~P. Townsend, and Kelly~R. Sutherland.
\newblock Cool your jets: biological jet propulsion in marine invertebrates.
\newblock 224(12):jeb222083.

\bibitem{larsonCostsTransportScyphomedusa1987}
R.~J. Larson.
\newblock Costs of transport for the scyphomedusa \textit{Stomolophus meleagris} l. agassiz.
\newblock 65(11):2690--2695.

\bibitem{demontMechanicsJetPropulsion1988}
M.~Edwin Demont and John~M. Gosline.
\newblock Mechanics of jet propulsion in the hydromedusan jellyfish, polyorchis penicillatus: I. mechanical properties of the locomotor structure.
\newblock 134:313--332.
\newblock {ADS} Bibcode: 1988JExpB.134..313D.

\bibitem{megillModulusElasticityFibrillincontaining2005}
William~M. Megill, John~M. Gosline, and Robert~W. Blake.
\newblock The modulus of elasticity of fibrillin-containing elastic fibres in the mesoglea of the hydromedusa polyorchis penicillatus.
\newblock 208(20):3819--3834.

\bibitem{gambiniMicroMacrorheologyJellyfish2012}
Camille Gambini, Bérengère Abou, Alain Ponton, and Annemiek J.~M. Cornelissen.
\newblock Micro- and macrorheology of jellyfish extracellular matrix.
\newblock 102(1):1--9.

\bibitem{demontMechanicsJetPropulsion1988a}
M.~Edwin Demont and John~M. Gosline.
\newblock Mechanics of jet propulsion in the hydromedusan jellyfish, polyorchis penicillatus:{II}. energetics of the jet cycle.
\newblock 134(1):333--345.

\bibitem{xuFieldTestingBiohybrid2020}
Nicole~W. Xu, James~P. Townsend, John~H. Costello, Sean~P. Colin, Brad~J. Gemmell, and John~O. Dabiri.
\newblock Field testing of biohybrid robotic jellyfish to demonstrate enhanced swimming speeds.
\newblock 5(4).

\bibitem{rutledge-in-prep}
Kelsi~M. Rutledge, S.~P. Colin, Noa Yoder, Simon~R. Anuszczyk, Kelly~R. Sutherland, Brad~J. Gemmell, and John~O. Dabiri.
\newblock Biohybrid robotic jellyfish for swimming-enhanced vertical ocean profiling.

\bibitem{anuszczykIncreasingReliabilityVersatility2025}
Simon~R. Anuszczyk, Noa Yoder, John~H. Costello, John~O. Dabiri, Brad~J. Gemmell, Kelsi~M. Rutledge, and Sean~P. Colin.
\newblock Increasing the reliability and versatility of jellyfish biohybrid vehicles via species selection and rhopalia removal.
\newblock 10(12).

\bibitem{liberzonOpenPIVOpenpivpythonFixed2019}
Alex Liberzon, Davide Lasagna, Mathias Aubert, Pete Bachant, jakirkham, ranleu, tomerast, Theo Käufer, Joe Borg, Cameron Dallas, and Boyko Vodenicharski.
\newblock {OpenPIV}/openpiv-python: fixed windows conda-forge failure with encoding.

\bibitem{satterlie_neuronal_2002}
Richard~A Satterlie.
\newblock Neuronal control of swimming in jellyfish: a comparative story.
\newblock 80(10):1654--1669.

\end{thebibliography}
\end{document}